\documentclass[fleqn,usenatbib]{mnras}

\usepackage{newtxtext,newtxmath}
\usepackage[T1]{fontenc}
\usepackage{ae,aecompl}
\usepackage{url}
\usepackage{graphicx}
\usepackage{xspace}
\usepackage{amsfonts}
\usepackage{pifont}
\usepackage{textcomp}
\usepackage{float}
\usepackage{subfig}
\usepackage{multirow}
\usepackage{adjustbox}
\usepackage{textgreek}
\usepackage{pdflscape}
\usepackage{threeparttable}
\usepackage{booktabs}

\title{V\;Sge: Supersoft Source or Exotic Hot Binary? I. An X-Shooter campaign in the high state}

\author[P. Hakala et al.]{
Pasi Hakala$^{1}$\thanks{E-mail: pahakala@utu.fi}, Phil Charles$^{2,3,4}$ and Pablo Rodríguez-Gil$^{5,6}$ \\
$^{1}$Finnish Centre for Astronomy with ESO (FINCA), Quantum, University of Turku, FI-20014, Finland\\
$^{2}$ Department of Physics \& Astronomy, University of Southampton, Southampton SO17 1BJ, UK\\
$^{3}$Astrophysics, Department of Physics, University of Oxford, Keble Road, Oxford OX1 3RH, UK\\
$^{4}$ Department of Physics, University of the Free State, PO Box 339, Bloemfontein 9300, South Africa\\
$^{5}$ Instituto de Astrofísica de Canarias, E-38205 La Laguna, Tenerife, Spain\\
$^{6}$Departamento de Astrofísica, Universidad de La Laguna, E-38206 La Laguna, Tenerife, Spain
}

\begin{document}
\label{firstpage}
\pagerange{\pageref{firstpage}--\pageref{lastpage}}

\outer\def\gtae {$\buildrel {\lower3pt\hbox{$>$}} \over 
{\lower2pt\hbox{$\sim$}} $}
\outer\def\ltae {$\buildrel {\lower3pt\hbox{$<$}} \over 
{\lower2pt\hbox{$\sim$}} $}
\newcommand{\Msun}{$M_{\odot}$}
\newcommand{\lsun}{$L_{\odot}$}
\newcommand{\Rsun}{$R_{\odot}$}
\newcommand{\solar}{${\odot}$}
\newcommand{\kep}{\sl Kepler}
\newcommand{\ktwo}{\sl K2}
\newcommand{\tess}{\sl TESS}
\newcommand{\swift}{\it Swift}
\newcommand{\Porb}{P_{\rm orb}}
\newcommand{\nuorb}{\nu_{\rm orb}}
\newcommand{\eplus}{\epsilon_+}
\newcommand{\eminus}{\epsilon_-}
\newcommand{\cd}{{\rm\ c\ d^{-1}}}
\newcommand{\MdotL}{$\dot{M}_{\rm L1}$}
\newcommand{\Mdot}{$\dot{M}$}
\newcommand{\Mdsolar}{$\dot{M}_{\odot}$~yr$^{-1}$}
\newcommand{\Ldisk}{L_{\rm disk}}
\newcommand{\src}{V\;Sge}
\newcommand{\ergscm} {erg s$^{-1}$ cm$^{-2}$}
\newcommand{\ergs} {erg s$^{-1}$}
\newcommand{\rchi}{$\chi^{2}_{\nu}$}
\newcommand{\chisq}{$\chi^{2}$}
\newcommand{\pcmsq} {cm$^{-2}$}
\newcommand{\Ha}{H\textalpha}
\newcommand{\Hb}{H\textbeta}
\newcommand{\Hg}{H\textgamma}
\newcommand{\Hd}{H\textdelta}
\newcommand{\Ion}[2]{#1\,{\sc #2}}
\newcommand{\Line}[3]{#1\,{\sc #2}\;\textlambda#3}
\newcommand{\kms}{\mbox{$\mathrm{km~s^{-1}}$}}
\newcommand{\he}[1] {He\,{\sc #1}}
\newcommand{\hel}[2] {He\,{\sc #1}\;\textlambda#2}

\maketitle

\begin{abstract}
\src\ is a peculiar, highly luminous long-period (12.34\;h) binary star that can display a super-soft X-ray emitting component when in the faint phase of its $V \approx 10-13$\;mag variability range. Apparently undergoing Eddington-limited accretion from its more massive secondary, it is in a very rare, short-lived evolutionary phase towards the double degenerate channel. Its complex and highly variable optical emission features, from Balmer and \Ion{He}{ii} to high-ionisation lines, including strong fluorescence features, have been challenging to interpret, especially given the absence of any absorption lines associated with photospheric features from either stellar component. With the detailed properties of \src, especially the donor, still controversial, we undertook a VLT/X-Shooter campaign over three months in 2023, obtaining high S/N, high resolution spectra that revealed multiple components in both high- and low-ionisation lines. This allows us to track \src's principal emitting regions via Doppler tomography, obtaining new insights into high accretion-rate dynamics.  In particular, we identify a stationary, double-peaked emission core which we interpret as a circumbinary ring, analogous to SS\,433. This enables us to derive limits on the system masses. Furthermore, we find very broad emission-line wings whose mean velocity can vary over hundreds of kilometres per second on timescales of decades, yet ``flip'' between states in $<\!1$\;week. We show that the super-soft X-ray source interpretation is able to account for these and other observational attributes significantly better than the hot binary model, concluding that \src\ could be one of the brightest known Galactic super-soft sources.
\end{abstract}


\section{Introduction and background to V\;Sge}
\label{sec:intro}
\src\ is a luminous ($L_\mathrm{bol} = 10^{37}$~\ergs) Galactic variable binary system that has been known since 1902 and has an orbital period of $P_\mathrm{orb} = 12.34$\;h. \cite{Herbig1965}, hereafter H65, identified its binary nature through both photometry and spectroscopy, concluding that it contained two evolved, hot stars of masses 0.7 and 2.8\;\Msun, the nature of which has remained controversial ever since.  Extensive modelling of the eclipsing light curve, assuming a pure binary model without any disc \citep[e.g.][]{Smak2001} indicated a binary inclination of 71\degr. The system exhibits high and low optical states (ranging over $V \approx 10.0 - 13.0$\;mag), which are possibly linked to changes in the accretion rate. A variety of models have been proposed and were most recently reviewed in \cite{Smak2001}, including that of a nova-like variable, a {\it s}uper{\it s}oft X-ray {\it s}ource (SSS), and a hot contact binary. Furthermore, it now has a \textit{Gaia} DR3-established distance of $3.02 \pm0.19$\;kpc \citep{2023A&A...674A...1G}, making it a very luminous object. Indeed, even in its faint state, \src\ 
has a luminosity 
$\sim\! 100$ times greater
than typical nova-like variables \citep{2020MNRAS.492L..40A}.

It was noted by \citet{Simon1999} that, in modelling the high state, the optical emission likely includes an additional, non-eclipsed component, as a consequence of the extremely high mass-loss rate from the donor, leading to the presence of circumbinary matter, as well as emission from an accretion disc wind \citep[see also top panel of fig.~3 in][]{Hachisu2003}. Occasionally, the high and low states seem to alternate with a cycle time of approximately one month \citep{Kato2004}. Various alternative models have been proposed, including hot detached binaries with colliding winds \citep{Lockley1999, Wood2000} or a hot binary submerged in a hot cocoon \citep{Smak2001}.

   \begin{figure*}
   \centering
   \includegraphics[width=0.85\textwidth]{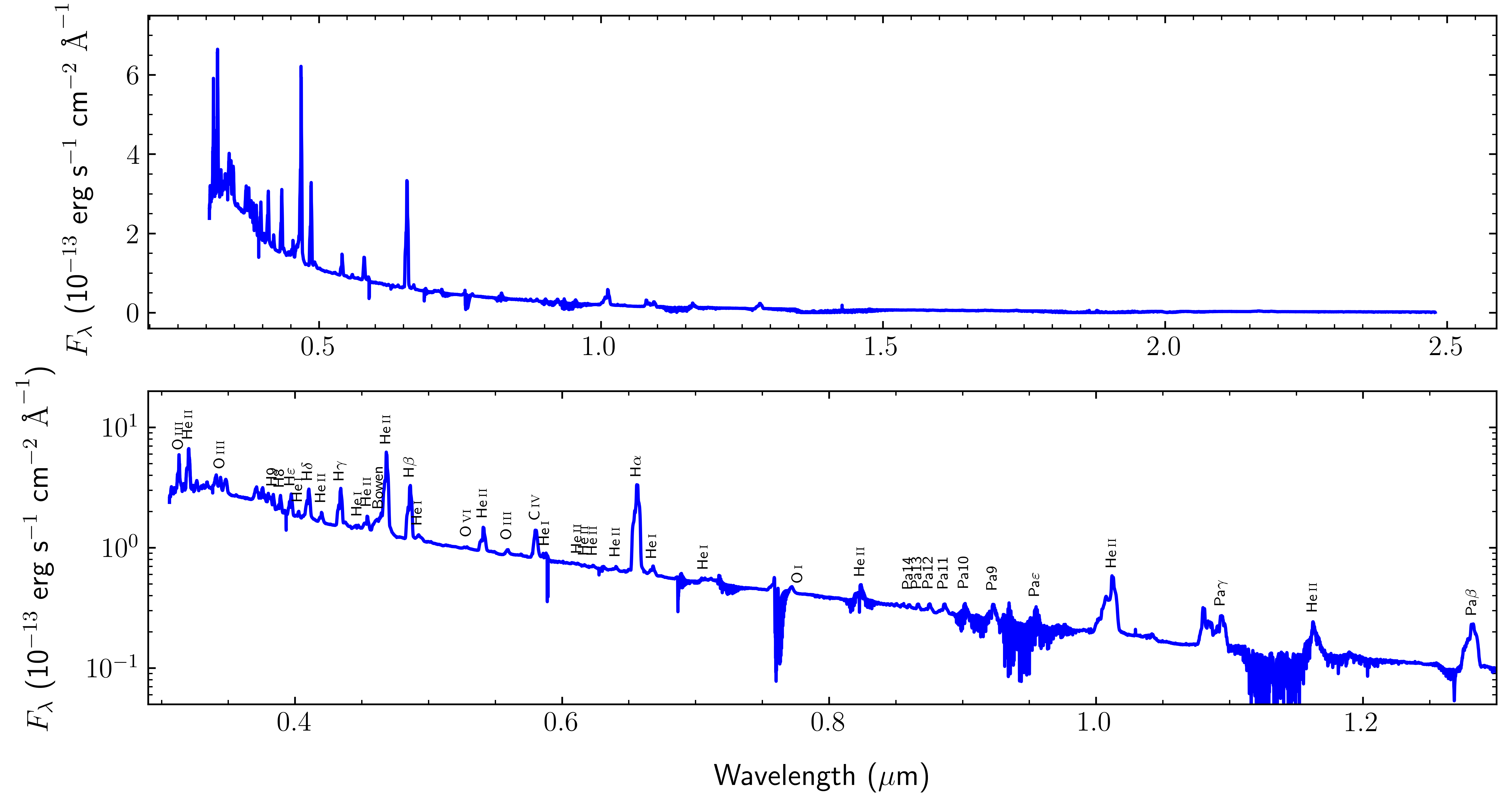}
    \caption{X-Shooter mean spectrum of \src\ in the high state on linear (upper) and logarithmic (lower) flux scales. The observed fluxes were corrected for interstellar reddening using $E(B-V) = 0.11$. An online interactive version of the top panel plot is available.\protect\footnotemark[1]} 
	\label{fig:avgspec}
    \end{figure*}
  \footnotetext[1]{\url{https://www.dropbox.com/scl/fi/odlhqugbm241diwxaxa1f/vsge_XSH_avg_spec.html?rlkey=25sl6n5w9qz7q8lf9dj9h7rrw&dl=1}}
  
The optical spectrum of \src\ is complex. The emission lines are not clearly double-peaked, which would typically indicate an accretion disc, but they occasionally exhibit P\;Cygni profiles, resembling the SW\;Sex behaviour in cataclysmic variables \citep{Thorstensen1991, Rodriguez-Gil2007}. Strong emission lines of \hel{ii}{4686} and the Balmer series, along with additional very high excitation lines (e.g. \Ion{O}{vi} and \Ion{N}{v}), are present in the spectrum, sometimes showing sharp, multiple components that vary with orbital phase (H65).  Both \he{ii} and \Hb\ vary over the orbit, and \cite{Diaz1999} attempted to Doppler map \citep{Marsh1988} the emission regions. However, this effort had limited success, likely due to the presence of significant vertical structure in the system. Such structure introduces velocity components perpendicular to the orbital plane, thereby violating a fundamental assumption of Doppler tomography. Consequently, the geometry underlying the complex and variable emission-line profiles remains poorly constrained.
  
Finally, the inferred high inverse mass ratio of the system, $q = M_2/M_1 \approx 3$, with $M_1$ and $M_2$ the masses of the accretor and the donor, primarily relies on the \Ion{O}{iii} fluorescent line radial velocity shifts observed by H65. These emission lines (\Line{O}{iii}{3133} and \textlambda3444) are reported to consist of two components that oscillate in anti-phase over the orbital period (see fig. 4 in H65). However, closer inspection of the radial velocity values in H65 reveals that the modulation is clearly {\it not} sinusoidal but instead shows constant radial velocities covering approximately one third of the orbital cycle. Furthermore, abrupt transitions occur from large positive to large negative velocities at, or near, phase 0.0, casting doubt on whether the two \Ion{O}{iii} components represent the intrinsic radial velocities of the binary components. In fact, we demonstrate here that the two components in the \Ion{O}{iii} lines correspond to the narrow double-peaked core emission found in all emission lines, with the primary distinction being the lack of high-velocity components in the \Ion{O}{iii} features.

\subsection{Evidence for the SSS Component in \src}
\label{sec:intro_SSS}

The earliest soft X-ray detections of \src\ were reported in \citet{1991ApJ...382..290E}, based on archival Einstein Observatory IPC images, and \citet{1996PASP..108...81H} (see their Appendix), using ROSAT observations of PU\;Vul, which fortuitously had \src\ in the field of view (albeit 31\arcmin\ off-axis).  These ROSAT data recorded an X-ray spectral hardness ratio indicative of ``very soft X-rays'', but no spectral modelling was undertaken.

More importantly, \cite{Greiner1998}, \cite{Patterson1998}, and \cite{Steiner1998} noted that \src\ had four key properties:
\begin{enumerate}
\item presence of \Ion{O}{vi} and \Ion{N}{v} emission;
\item a ratio of \he{ii}/\Hb\ emission $> 2$;
\item high $V$-band absolute magnitude, $M_V$;
\item a deep, wide primary eclipse,
\end{enumerate}
which were all very similar to the known SSS binaries, thereby providing strong circumstantial evidence for including \src\ among them. Classical SSSs are accretion-powered interacting binaries where a white dwarf (WD) primary accretes matter from, in most cases, a more massive, Roche-lobe-filling companion \citep{1997xisc.conf..435K}. Due to the inverted mass ratio, accretion in SSSs is a self-accelerating process, meaning that they represent a short-lived, but very violent phase in binary evolution. Consequently, the accretion rate in SSSs can be very high ($\sim 10^{-7}$~\Mdsolar), thereby enabling sustained thermonuclear burning of the accreted matter on the surface of the WD, and allowing SSSs to reach Eddington-limited X-ray luminosities, typically a thousand times greater than in normal cataclysmic variables. 

The X-ray emission of SSSs is characterised by a very low temperature ($\sim 10^5$\;K) and a very high luminosity ($L_{\rm X} > 10^{36}$~\ergs) for a WD compact object due to the high accretion rate \citep{vdH1992}. Few SSS candidates are known in the Galaxy because of their extremely soft X-ray spectrum and the high levels of extinction in the Galactic plane; accordingly, far more have been identified in the Magellanic Clouds. The optical emission of SSSs is primarily driven by X-ray reprocessing in the accretion disc surrounding the WD. \cite{Meyer-Hofmeister1997} modelled the optical light curves of SSSs, demonstrating that the accretion disc rim must be vertically extended, likely due to the impact of the high-rate accretion stream on the outer disc.

Furthermore, it had already been well established for the prototypical SSSs, CAL83 and RX\;J0513.9--6951, that the very soft X-ray component was visible only when the system was in a faint state \citep{Southwell1996}, a property already noted as analogous to that of the VY\;Scl CV-subclass \citep{Steiner1998}. Since state changes are unpredictable, \cite{Greiner1998} summarise all the ROSAT observations in their table~1, making it clear that only a single interval (May 1994) contained \src\ in a faint state and this yielded the highest X-ray count rate seen by ROSAT. Unfortunately, that was using the HRI detector, which nominally has no spectral capability. But \citeauthor{Greiner1998} exploited the onboard HRI PH-channel distributions of those data to show that they were completely different (contained in low energy channels only) compared to the bright state's harder X-ray data.  We re-examine these data later.

Unfortunately, \src\ has remained almost permanently in the high state in recent times, preventing further faint-state observations with the currently available soft X-ray facilities ({\it Swift}, {\it Einstein Probe}, {\it eRosita}). When it next enters a faint state, new observations should be undertaken without delay.

\subsection{A new campaign}

In spite of many extensive and multi-wavelength observing campaigns in the decades since H65, \src\ displays peculiarities that have defied explanation within standard interacting binary models.  While the photometric light curves have been well established, it has proved much harder to obtain comparable high quality spectroscopic coverage across the full orbital cycle because of its awkward 12.34-h orbital period.  Here, we bring a new approach to investigating the various emitting components in \src\ by exploiting the superb wavelength coverage (3000\;\AA\ -- 2.5\;$\mu$) and spectral resolution of X-Shooter spread over a $\approx\!4$-month period so as to obtain the highest S/N optical/NIR spectra of \src\ yet, and with uniform phase coverage, all in the high state.

   \begin{figure}
   \centering
   \includegraphics[width=0.98\columnwidth]{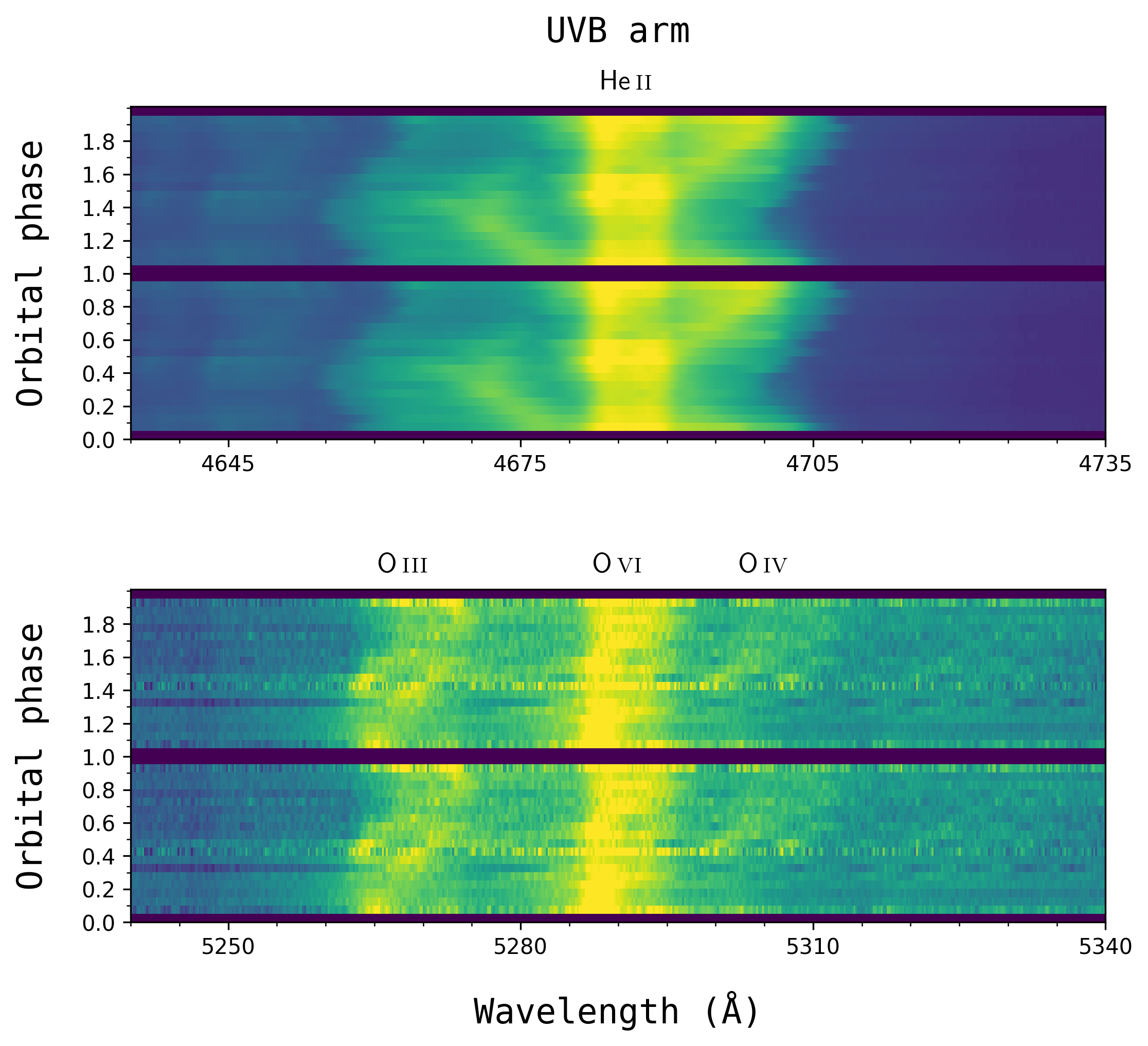}
    \caption{Trailed spectrograms showing the \hel{ii}{4686}  (upper) and \Line{O}{vi}{5290} (lower) emission lines. The spectra have been phase-binned into 20 orbital phase intervals and repeated once for display purposes. The dark horizontal stripes are bins containing no data. Note the systematic {\it blueshift} of the \Ion{He}{ii} broad line component. 
    } 
   \label{fig:HeII_4686__OVI_5290_trailed}
    \end{figure}

   \begin{figure*}
   \centering
   \includegraphics[width=0.98\textwidth]{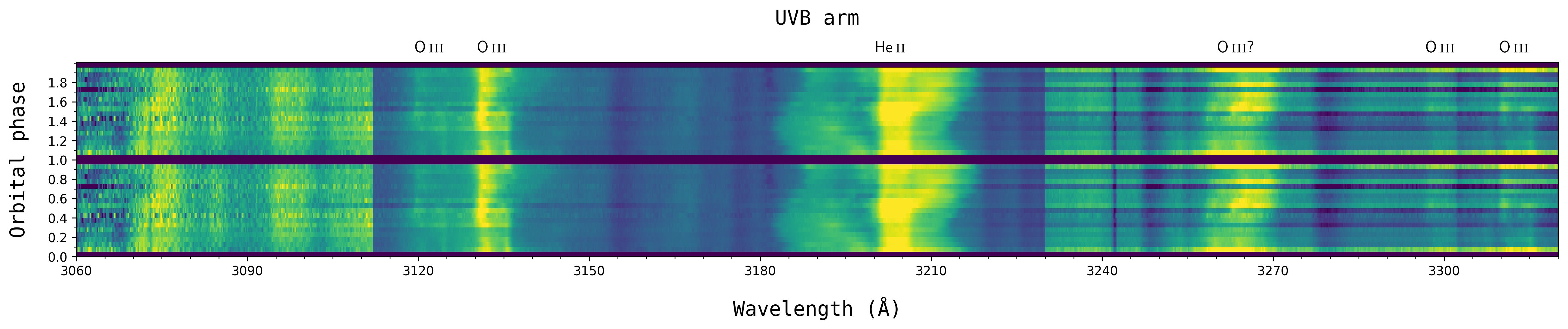}
    \caption{Trailed spectrogram showing the region around \hel{ii}{3203}. The spectra are phase-binned as described in Fig.~\ref{fig:HeII_4686__OVI_5290_trailed}.  Different contrast levels have been used in order to enhance the fainter lines, and one orbital cycle has been repeated for display purposes.
    } 
   \label{fig:HeII_3203_trailed}
    \end{figure*}

   \begin{figure}
   \centering
   \includegraphics[width=0.49\textwidth]{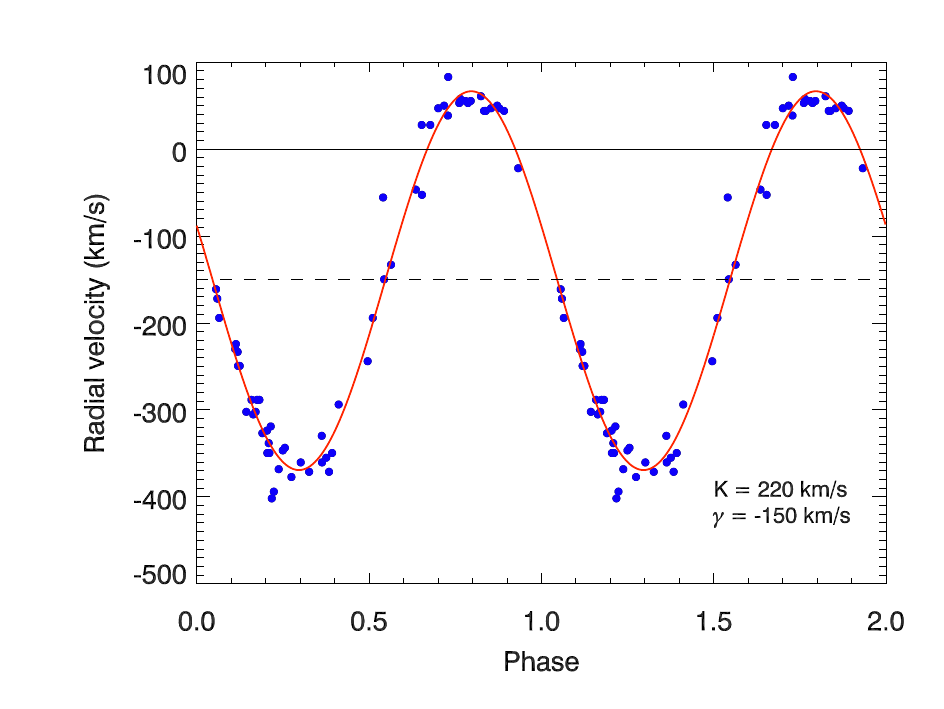}
    \caption{Radial velocity curves of the \hel{ii}{5411} emission lines (blue dots) derived from the double-Gaussian and wing-folding methods (Appendix~\ref{appendix:wings}). The best-fit sinusoid is shown in red. The radial velocity values for the individual spectra are plotted twice for clarity.
    } 
   \label{fig:HeII3203_HeII4686_RVCs}
    \end{figure}
   \begin{figure}
   \centering
   \hspace{-1cm}
   \includegraphics[width=0.53\textwidth]{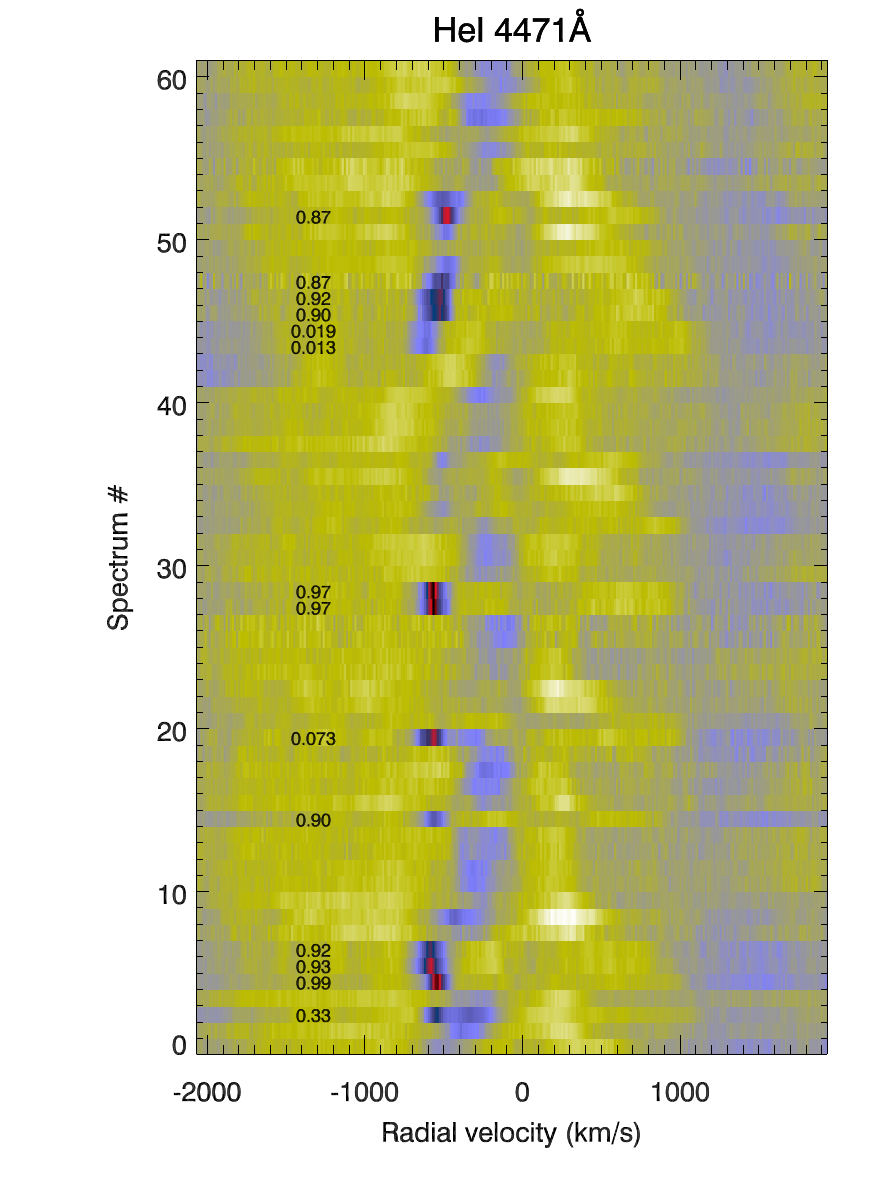}
 \caption{The individual spectra around \hel{i}{4471} in velocity space, ordered by spectrum number. The 61 spectra span an interval of $\approx\! 4$ months, with time running from bottom to
 top. Spectra exhibiting strong blue-shifted \Ion{He}{i} absorption have been marked with their respective orbital phase.  
} 
   \label{fig:HeI4471}
    \end{figure}

\section{Observations}

\src\ was the target of our X-Shooter campaign between Jun and Sep 2023 (proposal 111.24YF). 
X-Shooter is a versatile, medium-resolution  spectrograph located at the Very Large Telescope \citep[VLT;][]{Vernet2011}. Designed to cover a broad wavelength range, from the ultraviolet (3000\;\AA) to the near-infrared (25000\;\AA) in a single exposure, it provides simultaneous spectral coverage in three bands: UV-blue (UVB), visual (VIS), and near-infrared (NIR). With its medium spectral resolution ($R \approx 4000-18000$), it is well suited for both radial velocity and line profile studies.
Given the brightness of \src, it was observed as a filler target for less than ideal conditions, yielding approximately 60 good quality spectra over the 120-d observing period (60, 58, and 55 in the UVB, VIS and NIR bands, respectively). Even though these spectra were taken at random times, they adequately sample the 12.34-h orbital period of. We used 1\arcsec, 0.9\arcsec, and 0.9\arcsec\ slits in the UVB, VIS and NIR bands, respectively. This yielded spectral resolutions of approximately 5400, 8900, and 5600 in the same bands, respectively. No on-chip binning was applied. 

The X-Shooter spectra were reduced using the \texttt{ESOREFLEX} automated pipeline \citep{Freudling2013} together with the X-Shooter workflow.
This produces spectra corrected from the instrumental response; however, as our observations were conducted under non-optimal conditions, the flux calibrations are, in many cases, unreliable.  Nevertheless, intensive monitoring of \src\ by AAVSO showed that it remained in the high state throughout, with a mean magnitude of $V \approx 11$. The barycentric correction was also applied to the spectra (both in time and velocity).

\section{The X-Shooter spectra}

The overall spectrum of \src\ is shown in Fig.~\ref{fig:avgspec}. The reader can use the online zoom function in order to inspect any part of the spectrum in detail. The emission is dominated by a very blue continuum, accompanied by a plethora of strong emission lines with complex profiles. The strongest lines include the hydrogen Balmer series, various \Ion{He}{ii} lines, as well as \Ion{O}{iii} and \Ion{O}{\sc iv} lines (see Figs. \ref{fig:HeII_4686__OVI_5290_trailed}, \ref{fig:HeII_3203_trailed}, and Appendix \ref{appendix:trailed}). Notably, there are no clear absorption lines that could be associated with either stellar component. Detailed trailed spectra diagrams of all significant emission lines, from UVB to NIR, are presented in Appendix~\ref{appendix:trailed}.

All data were imported into \verb|MOLLY|\footnote{\url{https://cygnus.astro.warwick.ac.uk/phsaap/software/molly/html/INDEX.html}} to prepare them for trailed spectrograms and to measure the emission line radial velocity curves. 
 For simplicity, the orbital phase was computed using the linear eclipse ephemeris from \cite{Zang22}, with $P = 0.5141923$\;d, and a zero-phase time of $T_0 ({\rm HJD}) = 2460096.825$, derived from minima in AAVSO\footnote{The American Association of Variable Star Observers: \href{https://www.aavso.org}{aavso.org}} white-light ($CV$) and $V$-band light curves. The orbital phase shifts relative to the cubic ephemeris of \cite{Smak22} and the quadratic ephemeris of \cite{Zang22} are approximately 0.014 and 0.011 cycle, respectively, which we consider negligible.

%

   \begin{figure*}
   \centering
   \vspace{-0.0cm} 
\hspace{0cm}
\includegraphics[width=0.99\textwidth]{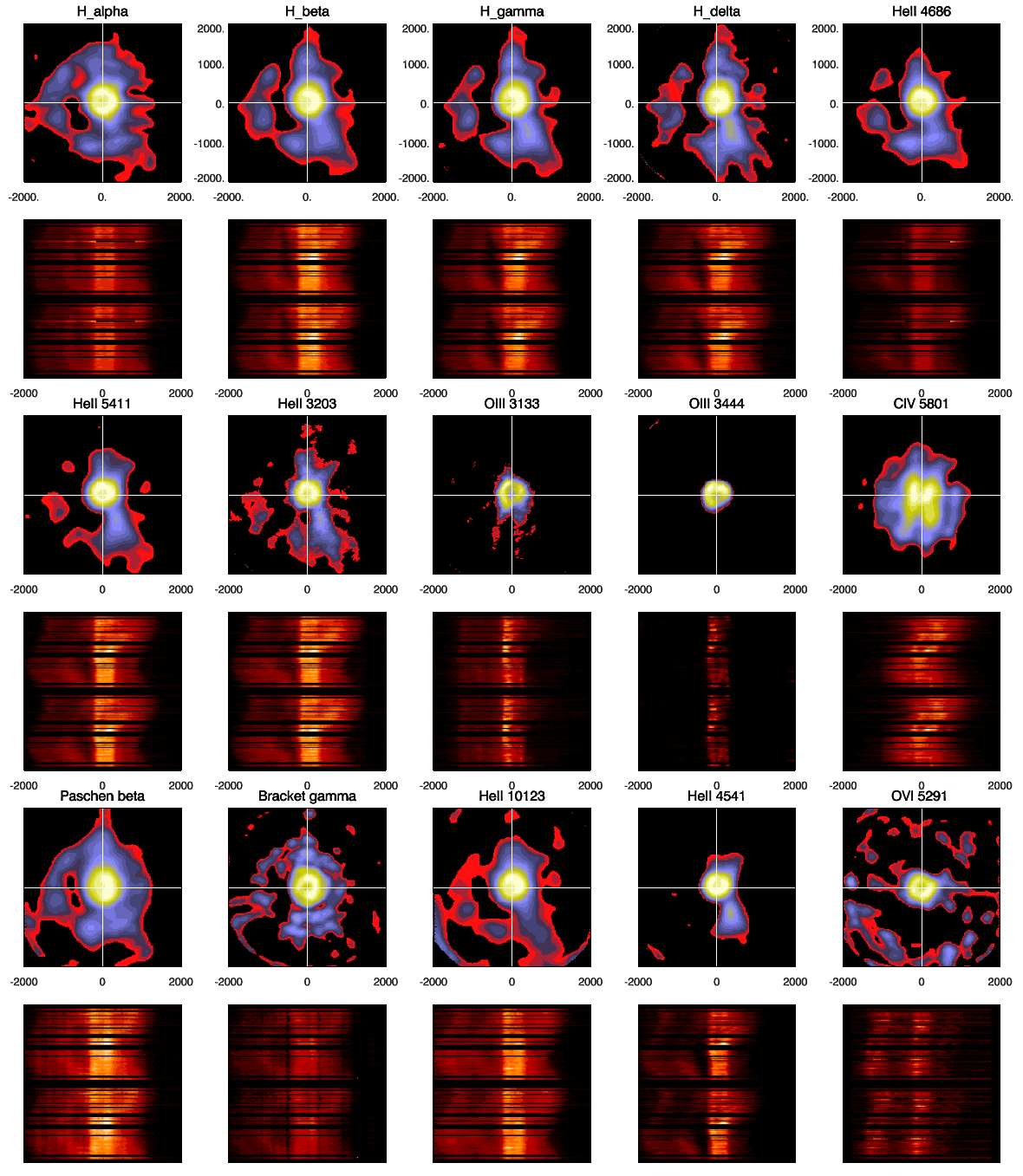}
\caption{Selected Doppler tomograms of various emission lines. For each line, we present the Doppler tomogram (top) and the corresponding trailed spectrogram (bottom). In the tomograms, the axes represent velocity coordinates $V_x$ and $V_y$ in \kms. In the trailed spectrograms, the $x$-axis indicates velocity in \kms, and the $y$-axis shows the orbital phase. For clarity, the trailed spectra are repeated over two orbital cycles.}
    
   \label{fig:dopmap}
    \end{figure*}

\subsection{Emission line profiles}
\subsubsection{Stationary double-peaked core (the ``tram lines'')}
The strongest emission lines in \src\ share a variety of common features showing similar variability (and non-variability) over the orbital period. Perhaps the most striking feature of almost all emission line profiles is the prominent, double-peaked narrow core emission with peaks at radial velocities of $\pm 150-200$\;\kms (Fig.~\ref{fig:HeII_4686__OVI_5290_trailed}, top).
This narrow line core is centred at the rest velocity and does {\it not} show any movement over the orbital period. This component, which has previously appeared as mostly flat-topped (due to spectral resolution limitations), is clearly double peaked in {\it all} lines. Notably, the velocities of the 
two peaks are close to those reported in H65 as the orbital velocity of the less
massive component in the system.  

\subsubsection{Doppler-shifted very broad component}
\label{sec:broad:comp}
Intriguingly, the Balmer and \Ion{He}{ii} lines also show a very broad emission component with a full width at zero intensity of $\mathrm{FWZI} = 3250$\;\kms, which appears to follow the motion of the presumed WD (which is eclipsed at phase 0.0). We have studied the dynamics of the broad emission component as a function of orbital phase employing the double-Gaussian method commonly used for modelling the wing motion of strong broad emission lines \citep{Schneider+Young1980, Shafter1983}.  We have also developed another similar approach, and chosen \Line{He}{ii}{5411} to carry out the analysis, as the \Line{He}{ii}{4686} blue wing is contaminated by the Bowen blend. The details of our new `wing-folding' method, and the double diagnostic diagrams are shown in Appendix~\ref{appendix:wings}. As a result, we find 
that the overall velocity shift of the broad component is $-\!150$\;km~s$^{-1}$ relative to
the \src\ systemic velocity and the double-peaked core component. Our analysis also reveals that the broad component exhibits a velocity modulation which follows the motion of the primary (see Fig.~\ref{fig:HeII3203_HeII4686_RVCs}), and implies $K_1\!=\!200\!-\!250$\;km~s$^{-1}$ for $i\!=\!90\degr$ and
$K_1\!=\!220\!-\!275$\;km~s$^{-1}$ for $i\!=\!65\degr$, which we presume as the lower limit for the inclination, given the partly eclipsing nature of the system. 
This strongly suggests that the broad component may carry a wind velocity component towards the observer in addition to the orbital motion. This will be discussed in detail later. 

The same emission lines also show evidence of a third, higher-velocity component that appears as an increasingly blue-shifted protrusion at phases 0.0--0.5 and
as a red-shifted protrusion at phases 0.5--1.0. This component reaches radial velocities of at least up to $\pm 1000$\;\kms, and its velocity phasing is consistent with that of a possible accretion flow {\it around} the WD. 
This is likely related to \src\ being Eddington-limited in the high state and this is producing an irradiation-driven outflow (or wind) from the inner disc, and possibly the WD surface.
In fact, at $M_V = -3.73$\;mag in the highest state, \src\ is intrinsically the most luminous of the known SSSs \citep{2003A&A...406..613S}.

\subsubsection{Blue-shifted absorption component}
The \Ion{He}{i} emission lines (at 3867, 4026, 4471, 5016, and 5876\;\AA) exhibit a fundamentally different profile from the Balmer and \Ion{He}{ii} lines. Each features a strong, phase-dependent, and blue-shifted absorption component with intriguing behaviour. The \Ion{He}{i} trailed spectra suggest a blue-shifted absorption component that steadily increases in velocity throughout the orbital cycle, reaching its maximum blue shift just before the phase gap covering the eclipse. However, a detailed inspection of  \Line{He}{i}{4471} (Fig.~\ref{fig:HeI4471}) in each individual spectrum reveals a strong, narrow absorption feature at $-500$ or $-600$\;\kms, which is predominantly present around phase 0.0. At other phases, the absorption  varies more in both width and central velocity. Moreover, its strength is significantly weaker at those phases. The combination of these effects results in the apparent phase-dependent blue-shifted behaviour observed in the \Ion{He}{i} trailed spectra. We have marked the spectra exhibiting strong blue-shifted absorption with their corresponding orbital phase in Fig.~\ref{fig:HeI4471}.
We note that this feature was previously observed by H65 and by \citet{2024PASJ...76.1002I}. The phasing of this blue-shifted absorption is particularly intriguing, as it suggests that cooler matter might be escaping the system from behind the donor star, i.e. through the binary's L2 point. We will explore this further in Section~\ref{sec:discuss}.  

\subsection{Doppler tomograms}
\label{sec:Doppler}

We have reservations about the application of Doppler tomography to a system such as \src, where significant vertical structures are likely present, leading to possible vertical velocity components and orbital-phase-dependent visibility issues with parts of the accretion disc, assuming geometries such as that proposed by \cite{Hachisu2003}. Nevertheless, we have conducted Doppler tomography of selected strong emission lines, as their principal orbital behaviour appears stable over extended intervals.

For this, we used the \texttt{DOPMAP} package of \cite{1998astro.ph..6141S}, which employs a maximum entropy approach for Doppler mapping, enabling comparison of model trailed spectrograms with those derived from the data. \texttt{DOPMAP} proceeds by first fitting and subtracting the continuum around the chosen line before producing the maximum entropy solution for the Doppler map. We selected all Balmer lines, as well as \Ion{He}{ii} and \Ion{O}{iii}, and the resulting tomograms are shown in Fig.~\ref{fig:dopmap}. We used $\gamma$ = 0.0 km/s for our Doppler analysis. The dominant feature in almost all Doppler maps is a bright central ring around the velocity origin. This ring has a typical radius of 150--180\;\kms\ and manifests as the line core ``tram lines'' in all the trailed spectrograms. There is no indication of any radial velocity changes associated with the ``tram lines'' structure itself. 

Additionally, extra emission appears in the south-east quadrants of the maps (i.e. positive $V_x$, negative $V_y$), which
could be linked to matter that has undergone a slingshot around the WD---possibly a quasi-ballistic accretion stream above the disc due to Eddington-limited accretion occurring in the high state, as mentioned earlier.
Notably, the fluorescent \Ion{O}{iii} lines (at 3133 and 3444\,\AA) also exhibit the same central emission ring but are much weaker at higher velocities.

Finally, the \Ion{C}{iv} line at 5804~\AA\ shows markedly different behaviour compared to all other lines. It is the only one that does \textit{not} display clear, non-variable ``tram lines'' in its core. Moreover, its overall profile appears to follow the motion of the WD more clearly than any other line.

\section{Archival Soft X-ray Observations}

\begin{figure}
   \centering
   \hspace{1cm}
   \includegraphics[width=0.5\textwidth]{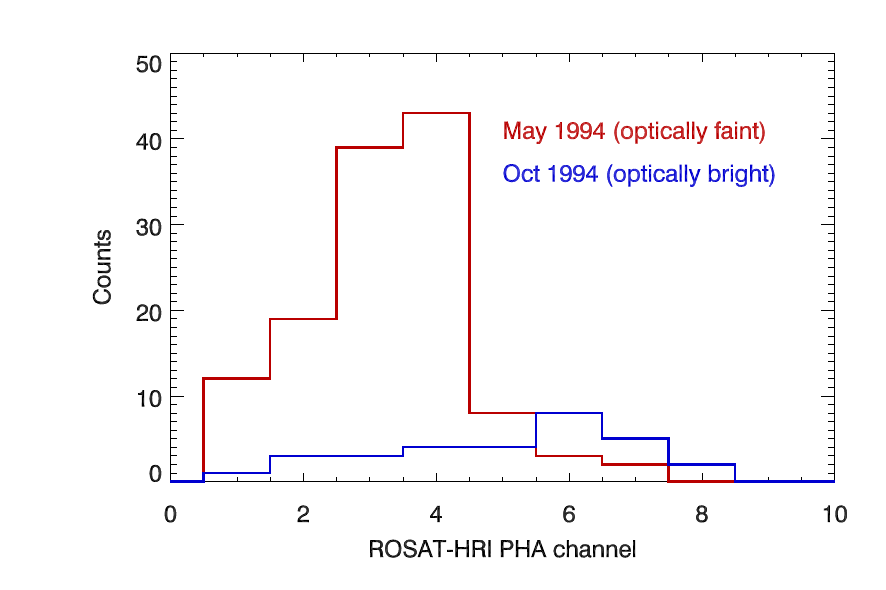}
    \caption{Soft X-ray pulse-height spectra of the ROSAT observations of \src\ with the HRI detector in the faint state of May 1994 (red) and the bright state of Oct 1994 (blue). PHA channel is proportional to X-ray energy, covering a range of $\approx\! 0.1\!-\!2.4$\;keV.
    } 
   \label{fig:ROSAT-HRI}
    \end{figure}

As summarised in Section~\ref{sec:intro}, there has only been one X-ray observation of \src\ in the faint state \citep{Greiner1998}, when the soft X-rays of a SSS component are typically visible. However, in a study of four VY\;Scl nova-like systems, \cite{Zemko2014} noted that they could detect no SSS component during any of their high or low state observations.  They go on to state that \src\ ``had not been actually observed as SSS'', which contradicts the work of \cite{Greiner1998}.  Accordingly, we have independently re-extracted the ROSAT source events for the two HRI observations in the bright and faint states, and in Fig.~\ref{fig:ROSAT-HRI} show the faint state (red) and bright state (blue) PH distributions.  To investigate the formal significance of this result, we performed a K-S test which gave a probability of 1.1 $\times$ 10$^{-6}$ that these two samples come from the same distribution.  This clearly demonstrates the presence of an SSS component at that time.

\section{Discussion}
\label{sec:discuss}


Our VLT/X-Shooter data provide an unprecedented spectral overview of \src, given the wide wavelength coverage and resolution. 
Despite---and perhaps because of---the quality and high S/N of these spectra, they are raising important questions regarding the nature of \src.   


\subsection{Do we know the basic binary parameters?}

Our current understanding of \src\ remains largely based on the initial spectroscopic study by H65, which continues to serve as the foundation for nearly all subsequent investigations.  The key properties identified by H65 are:
\begin{enumerate}
    \item two spectral line components varying in anti-phase, implying a mass ratio of $q=3.8$;
    \item component masses of $M_1 = 0.74\;M_\odot$ and $M_2 = 2.8\;M_\odot$.
\end{enumerate}

   \begin{figure*}
   \centering
   \vspace{-1cm}
   \includegraphics[width=0.7\textwidth,angle=90]{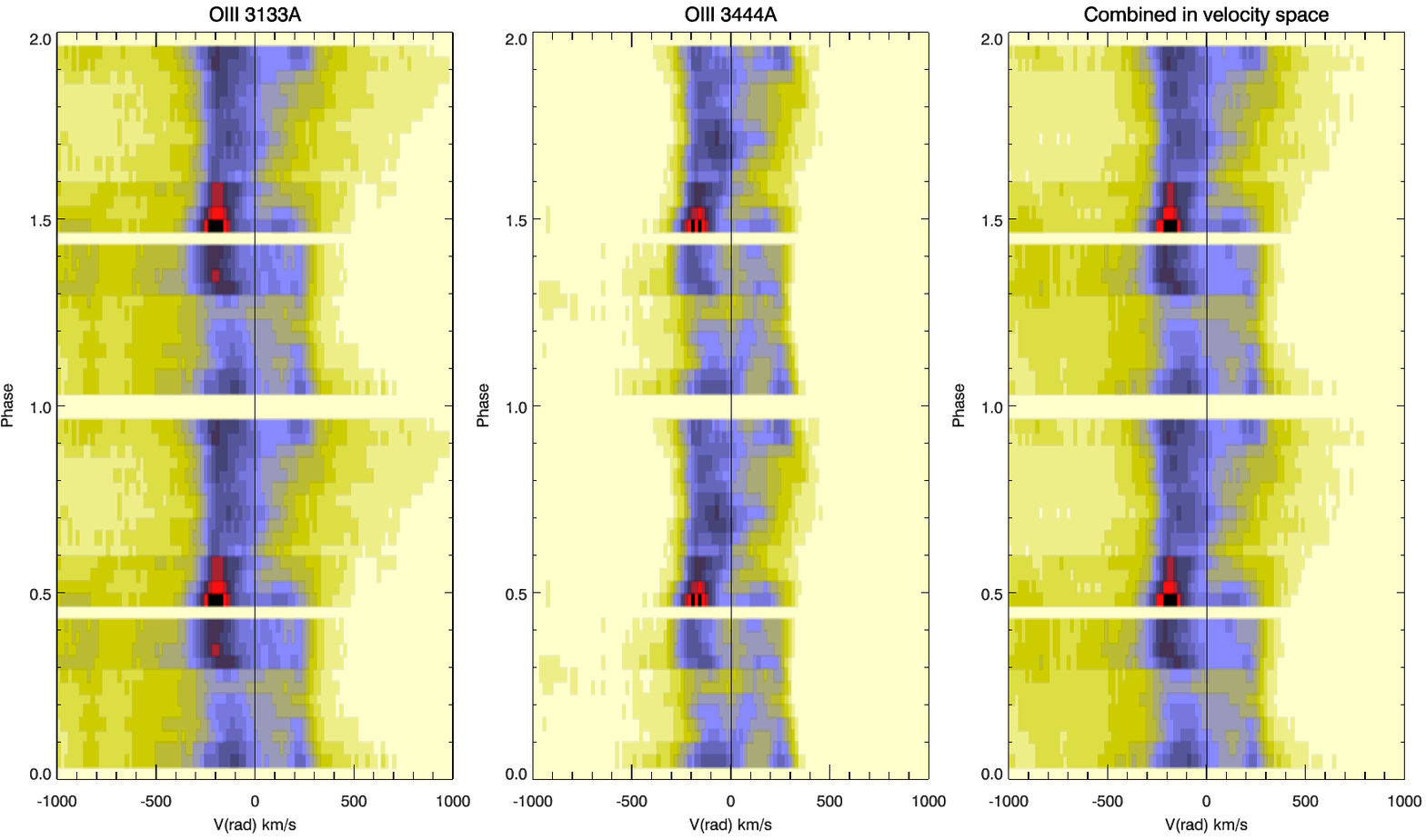}
   \vspace{-1cm}
    \caption{Trailed spectrograms of the \Line{O}{iii}{3133} and \textlambda3444 emission lines, previously used by H65 for determining the mass ratio of the system. The rightmost spectrogram shows the combined spectrogram of the \Line{O}{iii}{3133} and \textlambda3444 emission lines, earlier used for  lines in velocity space. There is no clear evidence of two spectral components varying in anti-phase. Note the inverse flux scale.}
    
   \label{fig:oiii_trails}
    \end{figure*}

Our primary concern regarding these properties is that, unlike most spectroscopic binary studies, \src\ is a purely {\it emission-line} object, with these estimates mainly derived from \Line{O}{iii}{3444} and \Line{O}{vi}{3811} measurements. We examined the \Line{O}{iii}{3133} and \textlambda3444 lines in our data set but found no clear evidence of two components moving in anti-phase (Fig.~\ref{fig:oiii_trails}). Moreover, as already mentioned in Section~\ref{sec:intro}, upon closer inspection, the original radial velocity curves of H65 do {\it not} exhibit a sinusoidal variation but instead show two components moving at constant velocities, with their relative strengths varying over $P_\mathrm{orb}$---much as is seen in our data. This raises the crucial question of whether we truly know the masses of the two stars in \src. We believe that the masses (and mass ratio) could well be quite different from what has been assumed in the past, as will be demonstrated next.
   \begin{figure}
   \centering
   \hspace{1cm}
   \includegraphics[width=0.5\textwidth]{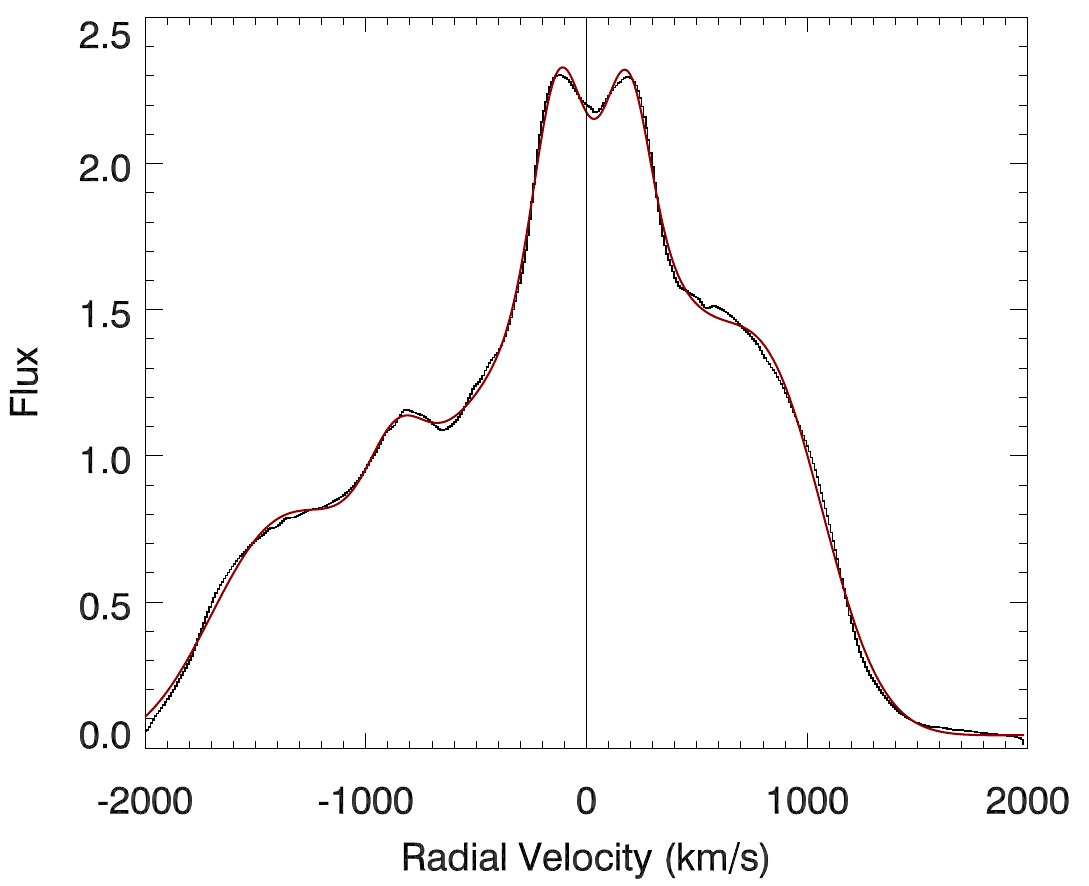}
    \caption{The mean \Ha\ line profile together with the best fitting model (red), consisting of seven Gaussian profiles used to estimate the circumbinary ring velocity.}
   \label{fig:mean_halpha}
    \end{figure}

   \begin{figure}
   \centering
   \hspace{1cm}
   \includegraphics[width=0.5\textwidth]{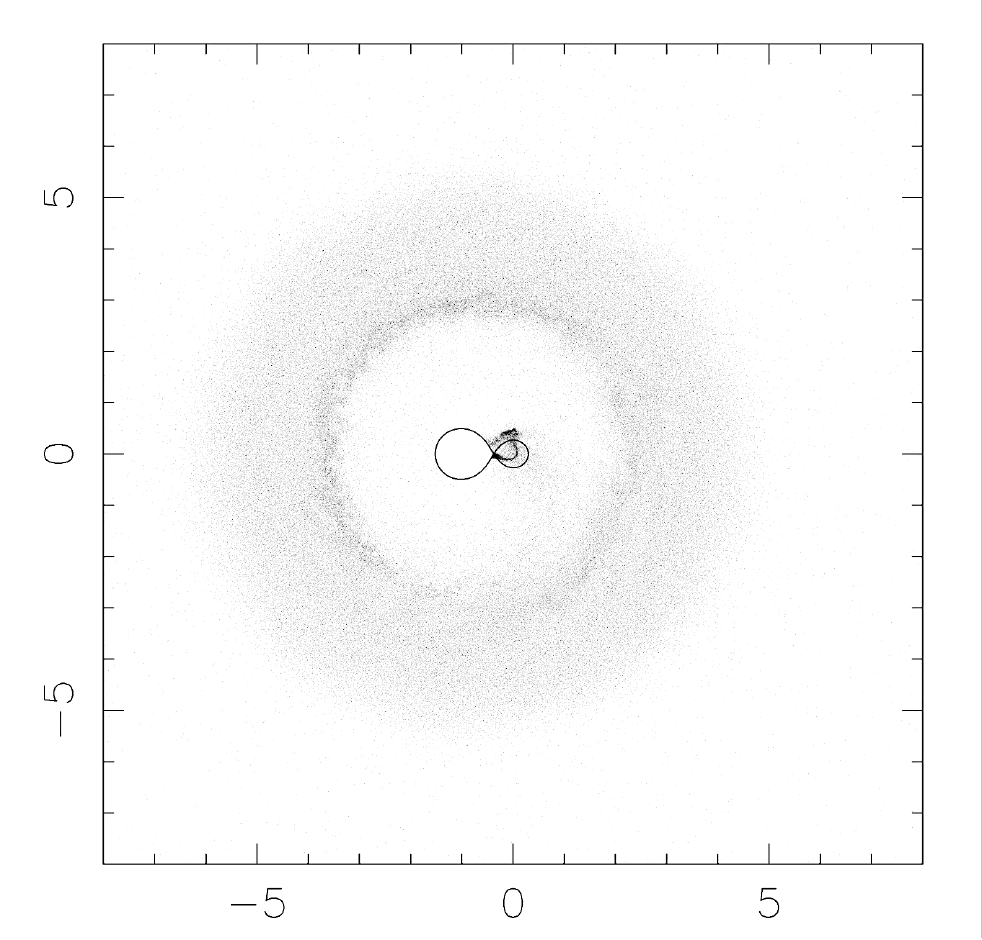}
   \includegraphics[width=0.51\textwidth]{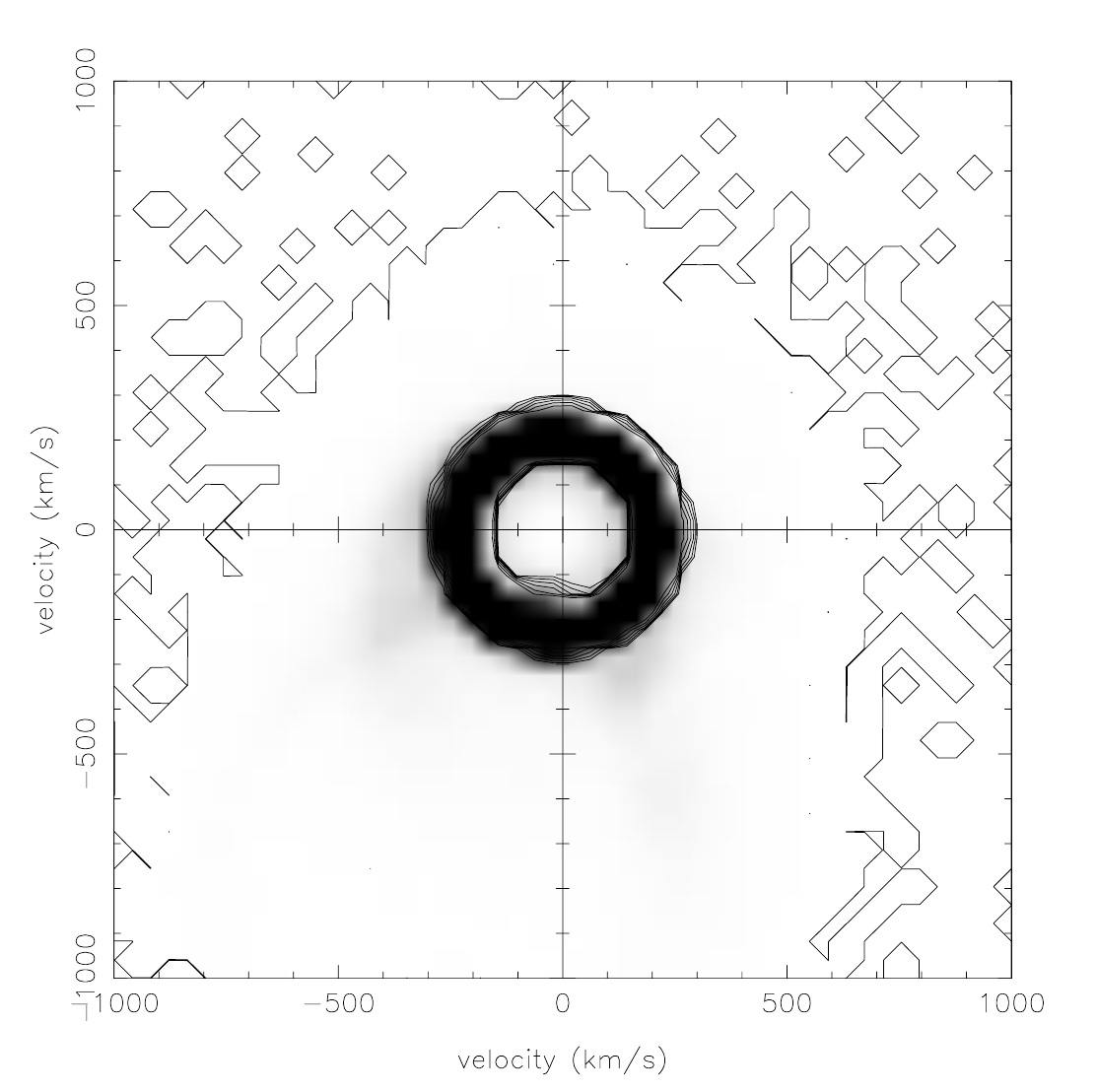}
    \caption{The resulting mass distribution from our simulation, showing the circumbinary ring (top panel; units are in binary separations) and the resulting Doppler tomogram (bottom panel).
    } 
   \label{fig:hydisc_points}
    \end{figure}

\subsection{Is there a circumbinary ring in \src?}

Our spectra cover the full X-Shooter range from 3000 to 24000\,\AA, and the plethora of detected emission lines includes the Balmer, Paschen, and Brackett series, numerous \Ion{He}{i} and \Ion{He}{ii} lines, as well as higher ionisation lines such as \Ion{O}{iii}, \Ion{O}{iv}, and \Ion{C}{iv}. A key feature revealed by our high spectral resolution is that most of these lines typically exhibit a narrow, double-peaked emission feature (the ``tram lines'' mentioned in Section~\ref{sec:Doppler}), with a peak separation of $\pm 150-180$\,\kms. More importantly, (a) this feature does \textit{not} vary with orbital phase, and (b) is centred at the systemic velocity, meaning that {\it these emission features do not follow the motion of either stellar component}. We have measured the orbital velocity of the circumbinary ring using two different methods. Firstly, we measure the radius of the central ring in the Doppler maps in eight different strong lines (Table~\ref{table1}) in both $X$- and $Y$-directions. This resulted in circumbinary ring velocities $\sim\!155$\;\kms. Secondly, we have fitted the mean \Ha\ profile with a model consisting of seven Gaussian profiles (Fig.~\ref{fig:mean_halpha}) applying the Differential Evolution (DE) global optimisation algorithm
\citep{1997JGOpt..11..341S}. We then bootstrapped the errors from this fit. This produced a circumbinary ring velocity of 159 $\pm$ 5\;\kms, compatible with the first method. This analysis also produced an estimate
for the systemic velocity, i.e. $\gamma = 28.6 \pm 1.9 $\;km~s$^{-1}$. We believe the only plausible explanation for this behaviour is that they originate from a circumbinary ring or disc of matter that has escaped the binary. 

   \begin{figure}
   \centering
   \hspace{0cm}
   \includegraphics[width=0.5\textwidth]{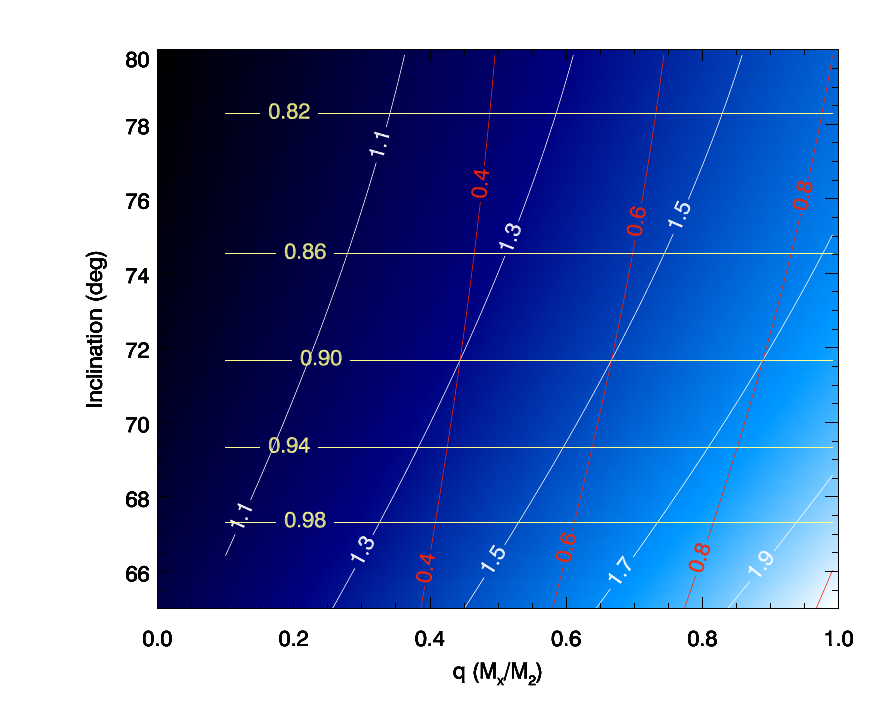}
    \caption{Possible system masses based on the measured circumbinary disc and primary star velocities as a function of $q$ and $i$. The total masses are shown with the colour map and white contours. The resulting primary masses ($M_\mathrm{X}$) are over-plotted with red contours, and the donor masses ($M_2$) with horizontal beige lines.}
    
   \label{fig:q_inc}
    \end{figure}

\begin{table}
\centering
  \caption{Estimated circumbinary ring orbital velocities$^a$ for different Doppler map emission lines.}
  \begin{tabular}{lcc}
     \toprule\noalign{\smallskip}
Line & $V_{\mathrm{circ},x}$ & $V_{\mathrm{circ},y}$  \\ 
   & (\kms) &  (\kms) \\
     \midrule\noalign{\smallskip}
\Ha & 181 & 193 \\
\Hb & 154 & 166 \\
\Hg & 154 & 166 \\
\Hd & 143 & 158 \\
\Line{He}{ii}{3203} & 132 & 132 \\
\Line{He}{ii}{4686} & 159 & 154 \\
\Line{He}{ii}{5411} & 153 & 145 \\
\Line{He}{ii}{10123} & 147 & 152 \\
     \midrule\noalign{\smallskip}
     $\bar V_\mathrm{circ}$: & $158 \pm 7$ & $153 \pm 5$ \\
     \bottomrule\noalign{\smallskip}
  \end{tabular}
       \\{\footnotesize $^a\,$The two columns relate to measuring the velocity profile of the central\\ map in the $x$- and $y$-directions. }
  \label{table1}
\end{table}

We have simulated the formation of such a ring using the binary magnetic accretion code \texttt{HYDISC} \citep{1993MNRAS.261..144K,1995MNRAS.275....9W}. To expel matter from the system, we introduced a ``magnetic propeller'' WD as the compact object, providing an additional ``kick'' to the matter when it falls on a ballistic trajectory past the WD. We then adjusted this ``kick'' to be as small as possible while still enabling at least some matter to escape the WD's Roche lobe. It is important to emphasise that we are not suggesting that \src\ contains a magnetic, spinning WD; rather, we use this simply as a means of expelling matter from the system. In reality, the matter is more likely to be driven out by radiation pressure, resulting from the near-Eddington-limited accretion and/or steady nuclear burning on the WD surface that we believe must be occurring.\footnote{Detailed simulations of this scenario are beyond the scope of this paper.}

For the simulation, we used $M_1=0.9$\,\Msun\ and $q=3.5$. We also experimented with different masses and mass ratios to assess the sensitivity of the simulation results to these parameters. A second simulation was carried out with $M_1=1.25$\,\Msun\ 
and $q=1$, to account for potential uncertainties in the radial velocity curves of H65. In both cases, the simulations produced a circumbinary ring  with typical velocities of $\simeq 200$\;\kms (Fig.~\ref{fig:hydisc_points}), and the results did not differ significantly between them.

Our simulations show that, as matter escapes from the WD's Roche lobe, it forms a ring or rings around the binary, with a radius of roughly 2--4 binary separations. The orbital velocity of the matter in such a circumbinary ring is $\approx\!200$\;\kms, somewhat larger, but broadly compatible with the double-peak separation we measure in most emission lines. 

It is interesting to note that the existence of a circumbinary ring has been proposed previously, based on both theoretical and observational studies. \citet{2002MNRAS.337..431P} simulated mass loss from luminous X-ray binaries due to the radiation pressure exerted on the donor star, resulting in mass loss via the outer Lagrangian point and the formation of a circumbinary ``excretion'' disc. \citet{2008ApJ...678L..47B}, on the other hand, observed that the emission line profiles in SS\,433 contain two narrow emission components that also do not vary with orbital phase. Their interpretation is similar to ours, i.e. they are likely caused by the presence of a circumbinary disc or ring. 

Following the approach taken by \citet{2008ApJ...678L..47B}, we can also obtain an 
independent measure for the total mass of the system as a function of the mass ratio, $q$. Assuming that the circumbinary ring radius is close to the 
innermost stable orbit around the binary, we can further estimate the radius of the circumbinary 
ring to be $R_\mathrm{c}$ $\sim F\times a$ ($a$ being the binary separation), where $F=2.2\!-\!2.3$ \citep{1999AJ....117..621H}. Then, we can use the equation
(with $q=M_\mathrm{X}/M_2$; $M_\mathrm{X} = M_1$):

\begin{equation}
    M_\mathrm{total}=\frac{V_\mathrm{c}^2 F V_x P_x}{2\pi G}(1+q)~,
    \label{eqn1}
\end{equation}

\noindent
where, $V_c$, $V_x$, and $P_x$ refer to the circumbinary ring orbital velocity, the primary orbital velocity and the binary orbital period, to derive an estimate for the total system mass, since we have the measured velocities. To do this, we have adopted a range of values for the inclination (65\degr\ to 80\degr, since the source is partially eclipsing) and  $q$ = $M_\mathrm{X}/M_2$ = 0.1 to 1.0. If we further adopt $F=2.3$ and assume that our maximum $V_x \sin(i)$ = 250\,\kms\ corresponds to the $K_1$ velocity of the primary, we can compute a grid of possible system masses.  Combining these with the $q$ values, we can also compute a similar grid for the primary and secondary masses, $M_\mathrm{X}$ and $M_2$ (Fig.~\ref{fig:q_inc}).

Perhaps the most striking result is that the total mass of the system is constrained to be below
2.1\;$M_\odot$ for all the $q$ values $<\!1$, with this maximum mass being achieved for $q\!=\!1$. Furthermore, the primary mass is constrained
to be below $\sim$ 1.0$\;M_\odot$, confirming a WD primary, as is generally assumed. Similarly, the mass of the secondary, $M_2$, is constrained to be within 0.8--1.0\;$M_\odot$ for the 65\degr--80\degr\ inclination range. Given that the system's intrinsic luminosity
is at least 2\;dex larger than in any other CV, we have not considered a possibility where the donor star would
be less massive (i.e. a normal CV scenario is excluded and the system is assumed to be in the thermally unstable self-accelerating state of mass transfer). Now, given Eq.~\eqref{eqn1}, we want to consider how strict these derived mass upper limits are. The total mass is quadratically dependent on $V_\mathrm{c}$, but $V_\mathrm{c}$ is well constrained by the data, as is $P_x$. This leaves two possibilities open: perhaps the circumbinary ring is not located at the innermost stable orbit around the binary (i.e. $F\!\gg\!2.3$) or $V_x$ does not reflect the true velocity of the primary. Unfortunately, we cannot answer these questions based on our current data alone. 


\subsection{Long-term motion of the very broad component}

\begin{figure}
   \centering
   \hspace{1cm}
   \includegraphics[width=0.5\textwidth]{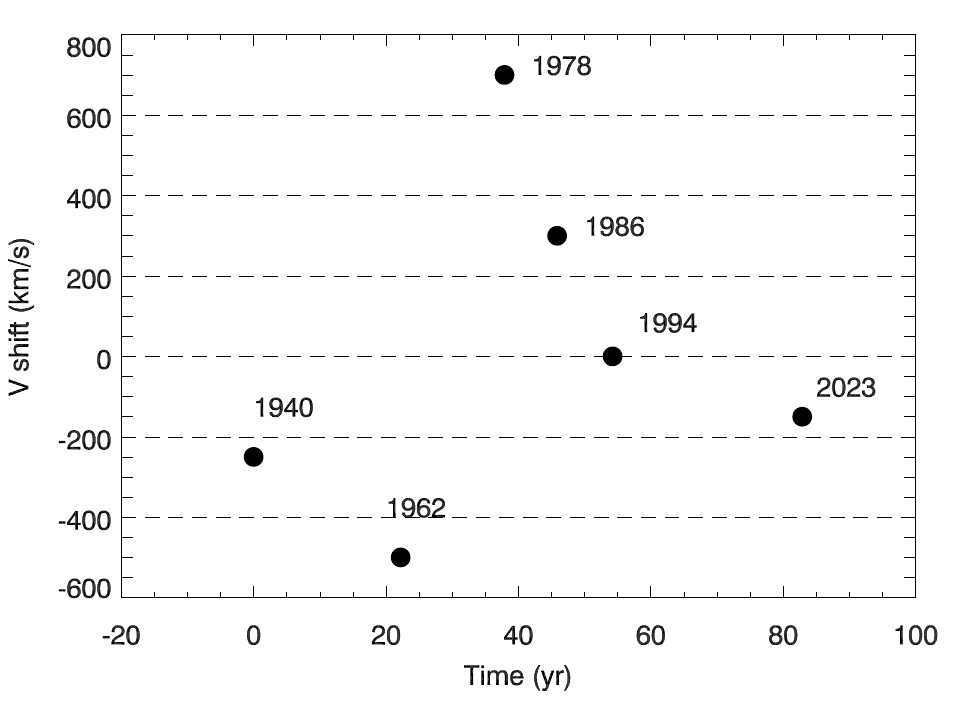}
    \caption{Historical velocity shifts since 1940 (time 0) of the centre of the broad emission line component \citep[from][and this work]{1943ApJ....97..412E, Herbig1965, 1986ApJ...306..618K, 1997AJ....113..787R, 1998AJ....115.2566G}. See also \citet{2024PASJ...76.1002I} for more recent, higher-cadence data.
    } 
   \label{fig:vshifts}
    \end{figure}

Another puzzling feature in the line profiles is the overall velocity shift of the (very) broad emission wings in strong emission lines, such as \Ha\ and \Line{He}{ii}{4686}. Our data show that, while the broad feature moves with orbital phase in contiguous data, its systemic velocity is measured at $-150$\;\kms. However, \citet{1998AJ....115.2566G} detected a systemic velocity of $+283\pm5$\;\kms\ in their \Ha\ study, while an earlier International Ultraviolet Explorer (IUE) FUV spectrum displayed a value of $+700\!\pm\!200$\;\kms\ \citep{1986ApJ...306..618K} in \Ion{He}{ii}, \Ion{C}{iv}, and \Ion{N}{v}. We then undertook a literature search that revealed a range of systemic velocities for this broad component dating back to 1940 (\citealt{1943ApJ....97..412E}; H65). We present these in Fig.~\ref{fig:vshifts}: while no clear long-term periodicity is evident in this $\approx\!100$\;yr of data, distinct and highly significant changes have occurred throughout this period. 


To produce such a broad line profile, the highest velocities must originate close to the accreting WD, i.e. within the inner accretion disc, and we observe this feature displaying orbital motion within each cycle.  But how, then, do we explain these extraordinary movements on timescales of years, which produce large, persistently blue- or red-shifted systemic velocities, while the narrow double-peaked feature remains essentially fixed at the binary's true systemic velocity?  We propose a possible explanation in which, due to its extremely high luminosity, the inner disc generates a strong, high-velocity wind responsible for the broad emission component. This could account for the overall line profile; however, to produce the large systemic velocity shifts on a timescale of many years, the disc would need to be tilted and/or warped and to precess over such an interval.

Very recently, \citet{2024PASJ...76.1002I} published the results of 26 years of optical spectral monitoring of \src. They find, as we do, that the source shows episodes of both blue- and red-shifted broad emission components. They also find that, while these states can persist for years, transitions between the blue- and red-shifted states can occur within just 4--5 days, imposing stringent constraints on the underlying physical mechanism. They present the temporal evolution of the broad and narrow line components in their fig.~10. 

Upon closer inspection of that figure, we find that the distribution of spectral states is bimodal, with the time the source spends in a blue- or red-shifted state ranging from days to several years. This behaviour appears chaotic, exhibiting two distinct ``modes'' with swift transitions between them. Physically, this could be linked to variations in the warping or tilt of the inner accretion disc, occurring on a timescale of days. The same tilt can then persist for an extended interval (from days to years) before the next ``flipping'' event takes place. 

We note that \citet{1999MNRAS.308..207W} demonstrated through numerical simulations that inner accretion discs become warped and precess due to radiation, {\it and}, if the radiation is strong enough, the inner disc's tilting behaviour becomes chaotic. According to \citet{2022AcA....72...21S}, the mass transfer rate $\dot{M}_{2}$ could be as high as $-2.5\times10^{-5}$\;\Msun\;yr$^{-1}$, placing \src\ well above the Eddington limit for a 1-\Msun\ WD. It is therefore plausible that chaotic changes in the inner disc inclination are the primary driver of the bimodal velocity shifts observed in the broad component of the emission lines. We should point out though, that the stability criteria of the inner accretion disc depend strongly on the accretion efficiency and the $\alpha$ parameter for the disc viscosity, as well as the bolometric luminosity, which is not well constrained for \src\ .

\begin{table}
\centering
  \caption{Comparison of different properties of \src\ and suggested models}
  \begin{tabular}{lcc}
     \toprule\noalign{\smallskip}
Property & Supersoft & Hot   \\ 
         & X-ray source    & binary\\ 
     \midrule\noalign{\smallskip}
Supersoft X-rays & \checkmark & \\
High \& low states & \checkmark & \\
Strong Balmer lines & \checkmark & \\
Highly ionised O, N, \& C & \checkmark & \checkmark  \\
Complex line profiles & \checkmark & \checkmark  \\
Non-variable line cores & \checkmark? & \\
Rapidly changing eclipse depths & \checkmark & \\
Overall mean optical light curve shape & \checkmark & \checkmark  \\
Emission lines with $\mathrm{FWZI} > 3200$\;\kms\ & \checkmark & \checkmark? \\ 
Changing orbital period & \checkmark & \checkmark? \\
Broad component velocity shifts & \checkmark? & \\  
     \bottomrule\noalign{\smallskip}
  \end{tabular}
  \label{table2}
\end{table}

\subsection{Is \src\ truly an SSS?}

While Fig.~\ref{fig:ROSAT-HRI} demonstrates the presence of very soft X-rays during the ROSAT May 1994 observation of \src\ during a faint state, we do not have a precise knowledge of its actual luminosity.  This is partly because the ROSAT HRI was only intended to be used as an X-ray imager, and so detailed spectral modelling is not possible.  However, this was also noted by \cite{Greiner1998}, whose fig.~2 (top) gives the $L_\mathrm{X}$--$T$ contours for fitting their strongest ``intermediate'' state ``soft X-ray'' spectrum, which was obtained with the ROSAT PSPC detector.  At first sight, this appears to suggest only a low $L_\mathrm{X}$ value of $\sim\!10^{32}$\;erg\;s$^{-1}$ (we have corrected for the now known distance of 3.03\;kpc, whereas \citeauthor{Greiner1998} used 1\;kpc), far below that expected for an SSS component.  However, when including the effects of an intervening absorbing column (we used HEASARC's $N_\mathrm{H}$ tool to calculate $N_\mathrm{H} = 1.56 \times 10^{21}$\;cm$^{-2}$), then their spectra require much higher $L_\mathrm{X}$ values, exceeding $10^{36}$\;erg\;s$^{-1}$, in order for the SSS component (likely $T<5\times10^5$\;K) to penetrate this material.

\subsection{Classifying \src: where does it fit?}

Finally, we discuss the nature of \src\ and the implications of our new spectral dataset. For several decades now, two different models have been proposed to describe the observed behaviour of the system. The original model of H65 assumes a hot binary system, in which both components are hot stars ($T_\mathrm{WD}=44\,000$\;K and $T_\mathrm{donor}=22\,000$\;K). This was also advocated by \citet{Lockley1999} and \citet{2000MNRAS.313..789W}, who claim that the emission-line profiles can be reproduced by a model involving colliding winds from such stars. This was also strongly preferred by \citet{2022AcA....72...21S}.

The second proposed model for \src\ suggests that it is a high-inclination Galactic SSS \citep{Greiner1998}, i.e.  a WD accreting matter from a more massive companion star via (unstable) thermal-timescale Roche-lobe overflow, as introduced in Section~\ref{sec:intro_SSS} and modelled by \citet{Hachisu2003}. The resulting accretion rate is so high that it can sustain ``steady'' nuclear burning on the WD surface \citep{vdH1992}. In such cases, the optical emission would be dominated by the accretion disc rather than by either of the two stars. 

To compare the merits of both models, we have compiled a list of the observed properties of \src\ in Table~\ref{table2}, and assessed how well they align with each model. While the hot binary model can explain about half of the listed properties, it fails to account for many crucial observed features. In particular, it does {\it not} explain the system's variability on either short-term (light curve shape changes over days) or long-term (low vs. high state) timescales. Additionally, the mechanism by which such a hot binary would generate extremely strong \Ha\ emission remains unclear. \Ha\ is the  second strongest emission line in the system after \Line{He}{ii}{4686}. Finally, the presence of stationary, double-peaked narrow emission lines would still necessitate the formation of a disc or ring around the binary, the origin of which remains unexplained within the hot binary model.

\section{Conclusions}


Our X-Shooter spectroscopic observing campaign of \src\ has reaffirmed the complexity of this system, highlighting that fully understanding its nature remains a significant challenge. Based on the presence of narrow emission-line components during the high state, we argue that \src\ likely hosts a circumbinary ring or disc. Historical data, together with recent spectroscopic monitoring \citep{2024PASJ...76.1002I}, suggest that the long-term variations in the mean velocity of the very broad component are probably driven by the chaotic behaviour of the inner accretion disc under intense irradiation. This interpretation aligns well with simulations by \citet{1999MNRAS.308..207W}.

We also raise doubts about the canonical radial velocity curves presented by H65, and consequently question the mass ratio and component masses derived from them. Our own dynamical analysis indicates that \src\ is likely a low-mass system, with a total mass less than $2.1\,M_\odot$. To test this hypothesis, high-resolution spectroscopic observations, particularly during the system’s faint state, are essential. Finally, from the observational properties summarised in Table~\ref{table2}, we conclude that \src\ is driven by a super-soft X-ray source (SSS) component, which appears to be the most luminous of its kind currently known in the Galaxy.

 

\section*{Acknowledgements}

Firstly, we thank the anonymous referee for a very constructive report with suggestions that we feel have greatly helped with the presentation of a large, complex dataset.
We would also like to thank Graham Wynn for access to his \texttt{HYDISC} magnetic accretion simulation code. PAC is grateful to Marina Orio for useful discussions on the SSS component of \src, and to Katherine Blundell regarding the circumbinary ring interpretation. PAC acknowledges the Leverhulme Trust for an Emeritus Fellowship, and UKRI for their ESO support.  PR-G acknowledges support by the Spanish Agencia Estatal de Investigación del Ministerio de Ciencia e Innovación (MCIN/AEI) and the European Regional Development Fund (ERDF) under grant PID2021--124879NB--I00.

\section*{Data Availability}

The X-shooter spectra are publicly available from the ESO data archive and the fully zoomable plot of the mean spectrum from the included link.
\vspace{4mm}



\bibliographystyle{mnras}
\bibliography{biblio} 


\appendix

\section{Trailed spectra plots}\label{appendix:trailed}

\begin{landscape}
\begin{figure}
\begin{center}
\includegraphics[width=0.85\linewidth]{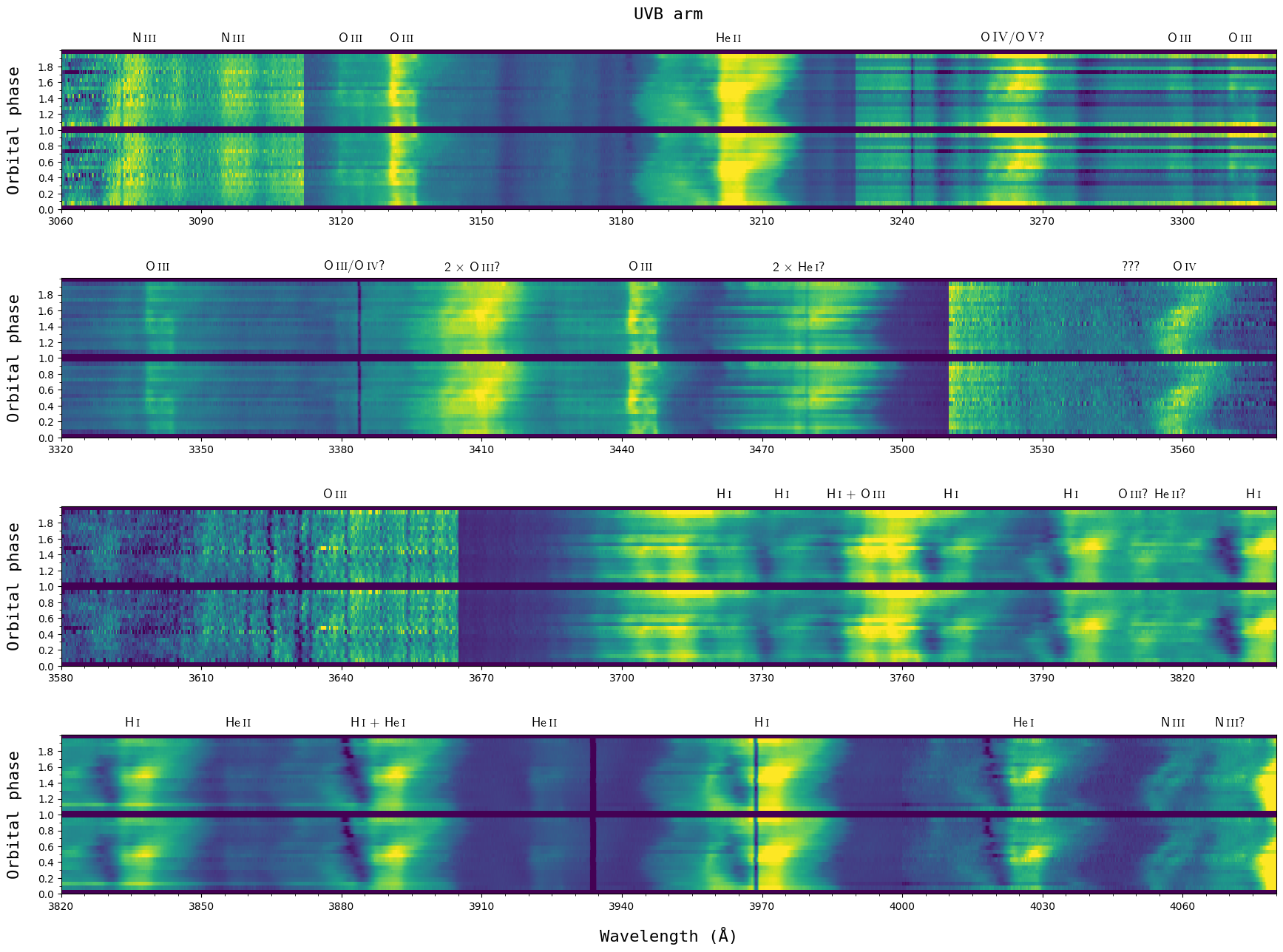}
\caption{X-Shooter UVB arm trailed spectra diagrams using all the data averaged into 20 orbital phase bins. The full cycle has been repeated once for clarity.}
\label{fig:trailed_UVB_01}
\end{center}
\end{figure}
\end{landscape}

\begin{landscape}
\renewcommand{\thefigure}{\thesection\arabic{figure} (cont.)}
\addtocounter{figure}{-1}
\begin{figure}
\begin{center}
\includegraphics[width=0.85\linewidth]{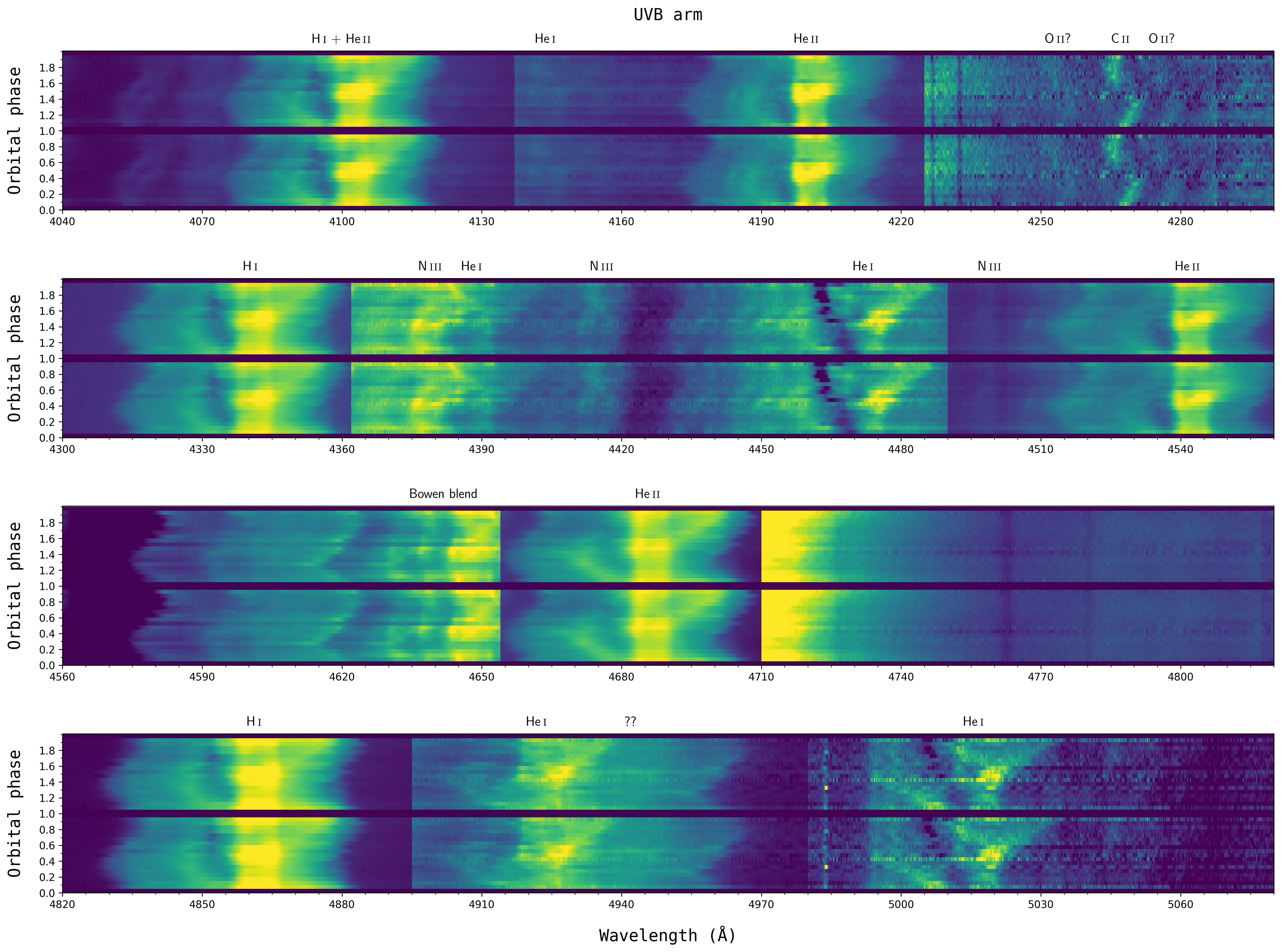}
\caption{}
\label{fig:trailed_UVB_02}
\end{center}
\end{figure}
\end{landscape}

\begin{landscape}
\renewcommand{\thefigure}{\thesection\arabic{figure} (cont.)}
\addtocounter{figure}{-1}
\begin{figure}
\begin{center}
\includegraphics[width=0.85\linewidth]{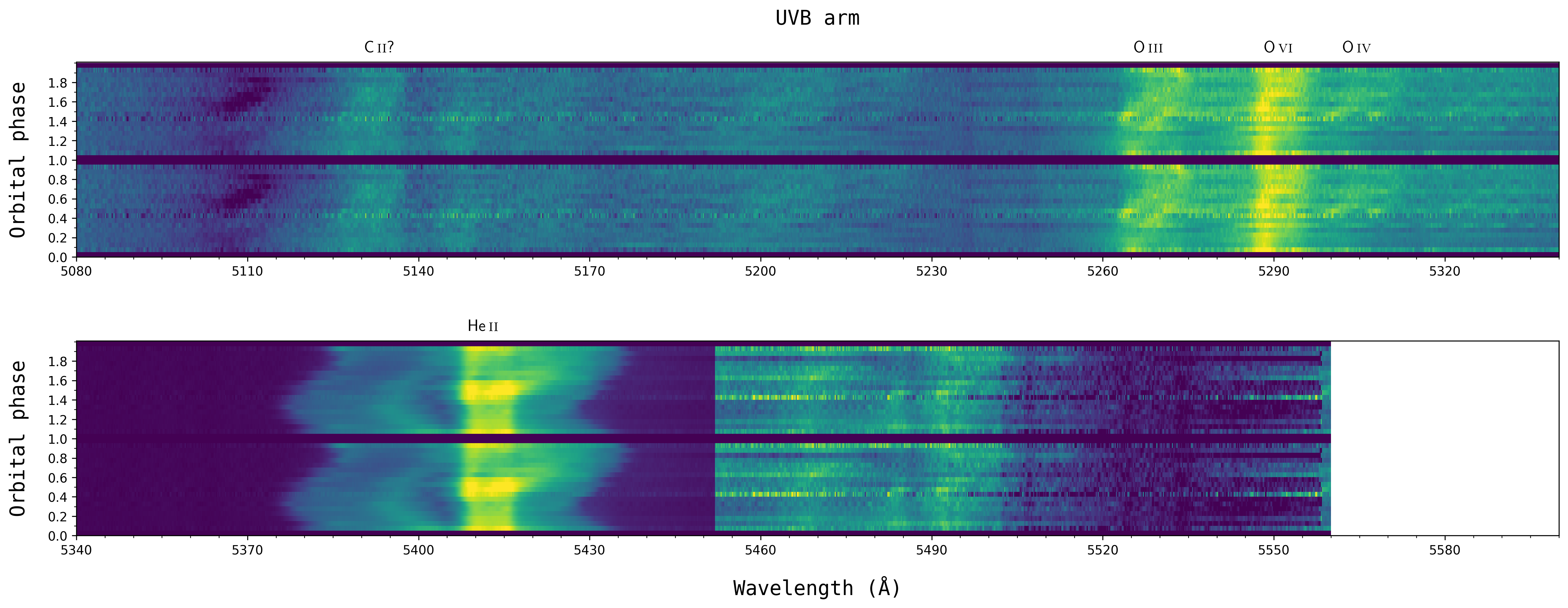}
\caption{}
\label{fig:trailed_UVB_03}
\end{center}
\end{figure}
\end{landscape}

\begin{landscape}
\begin{figure}
\begin{center}
\includegraphics[width=0.85\linewidth]{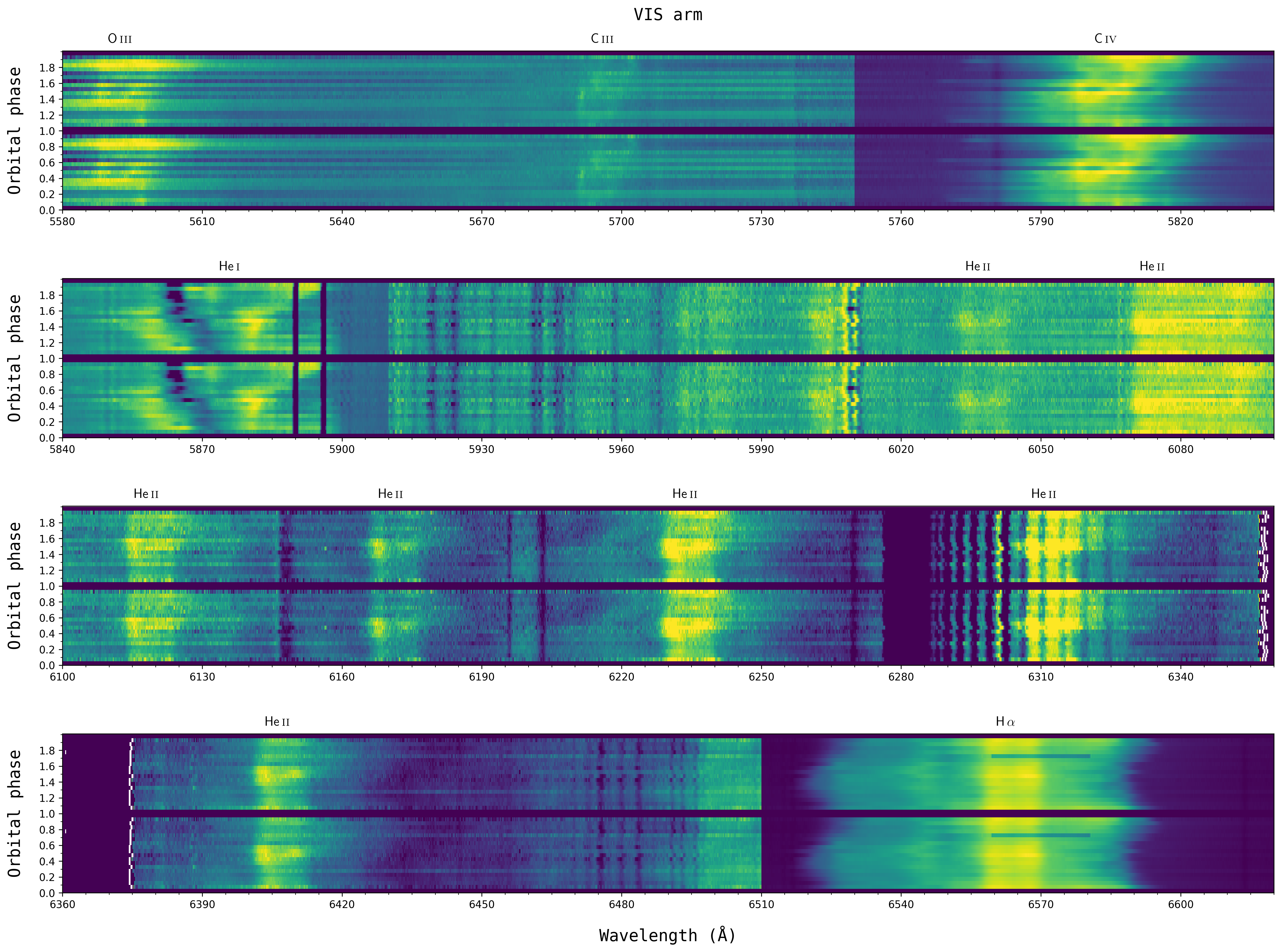}
\caption{X-Shooter VIS arm trailed spectra diagrams using all the data averaged into 20 orbital phase bins. The full cycle has been repeated once for clarity.}
\label{fig:trailed_VIS_01}
\end{center}
\end{figure}
\end{landscape}

\begin{landscape}
\renewcommand{\thefigure}{\thesection\arabic{figure} (cont.)}
\addtocounter{figure}{-1}
\begin{figure}
\begin{center}
\includegraphics[width=0.85\linewidth]{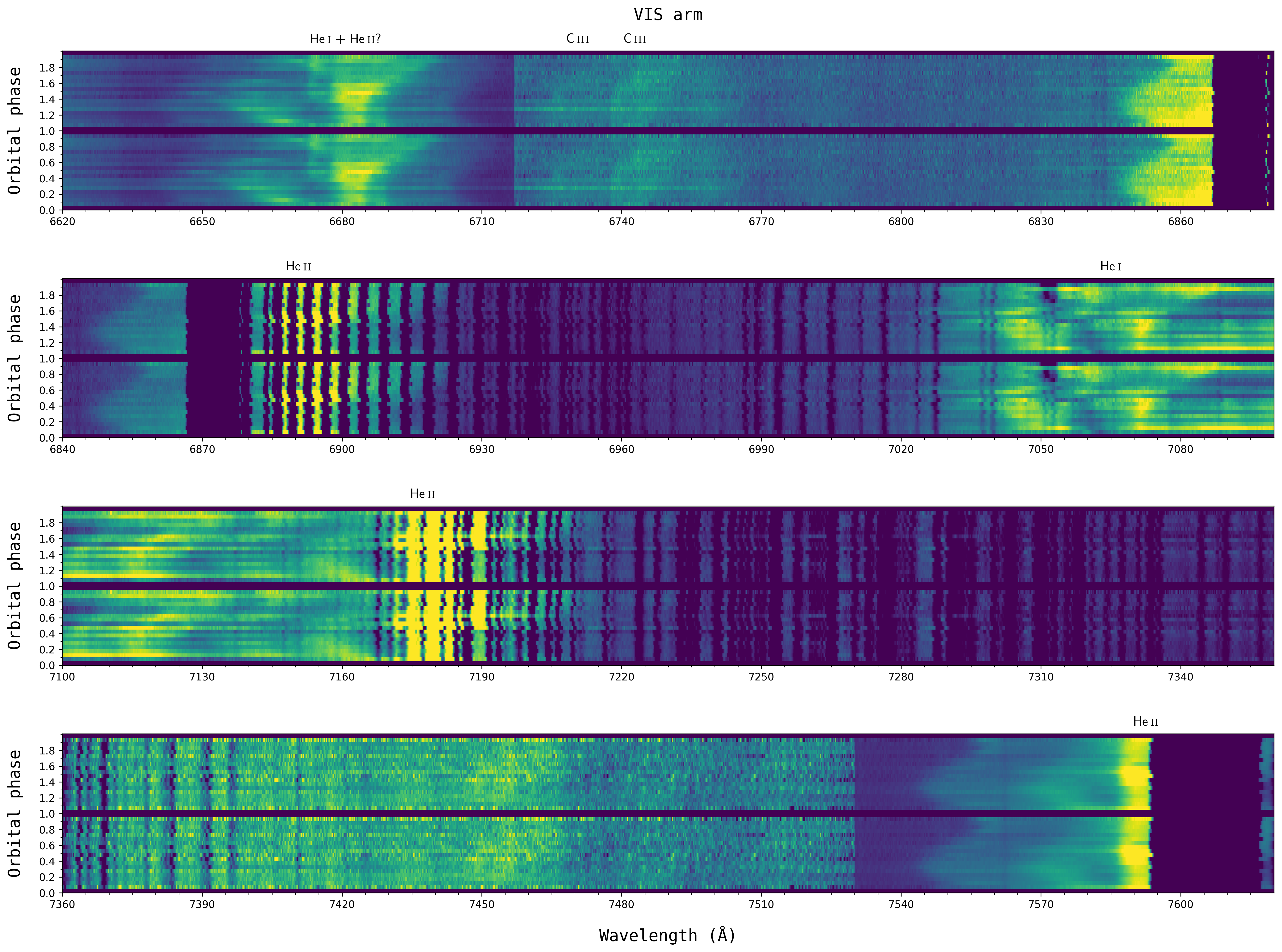}
\caption{}
\label{fig:trailed_VIS_02}
\end{center}
\end{figure}
\end{landscape}

\begin{landscape}
\renewcommand{\thefigure}{\thesection\arabic{figure} (cont.)}
\addtocounter{figure}{-1}
\begin{figure}
\begin{center}
\includegraphics[width=0.85\linewidth]{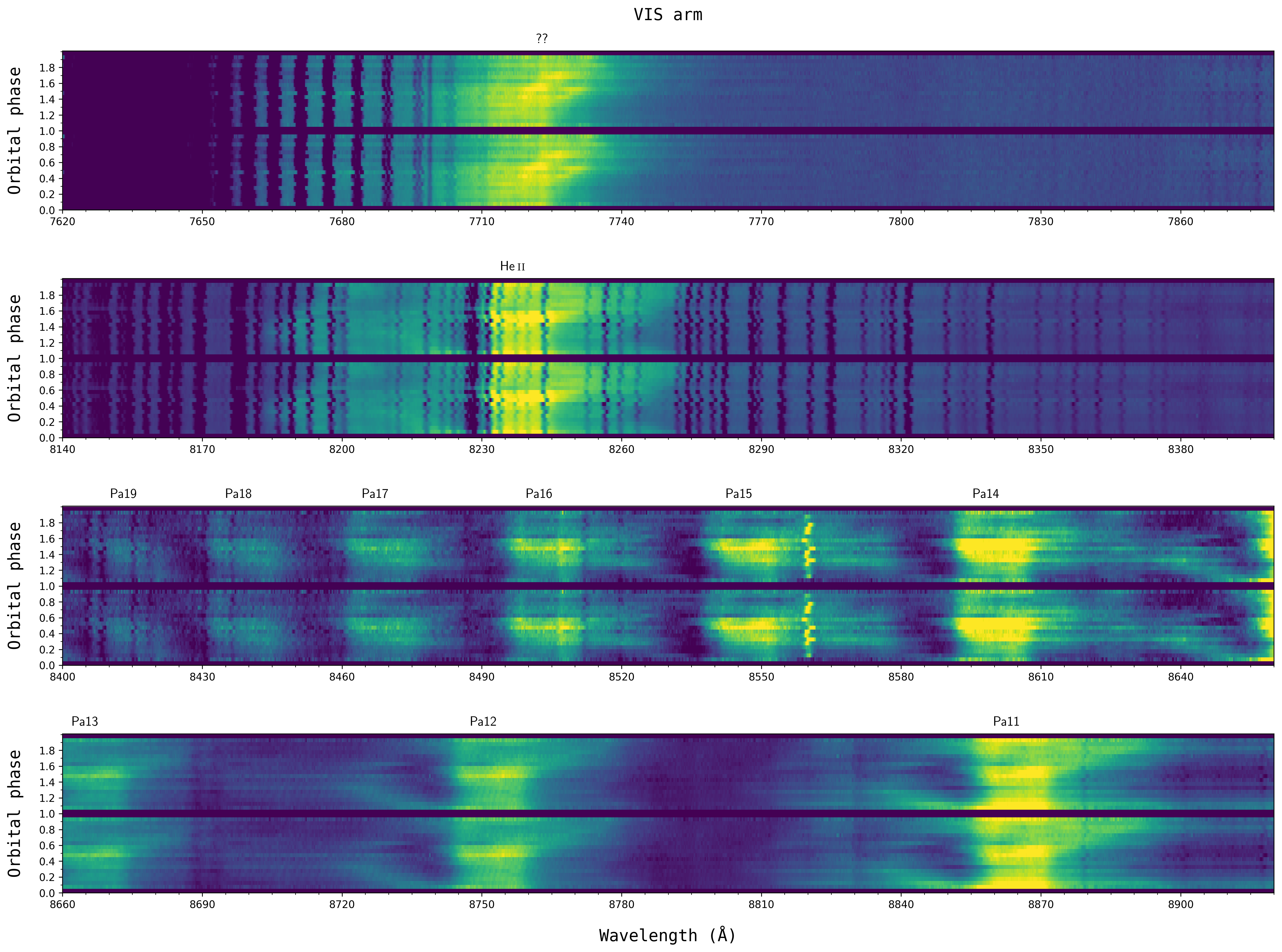}
\caption{}
\label{fig:trailed_VIS_03}
\end{center}
\end{figure}
\end{landscape}

\begin{landscape}
\renewcommand{\thefigure}{\thesection\arabic{figure} (cont.)}
\addtocounter{figure}{-1}
\begin{figure}
\begin{center}
\includegraphics[width=0.85\linewidth]{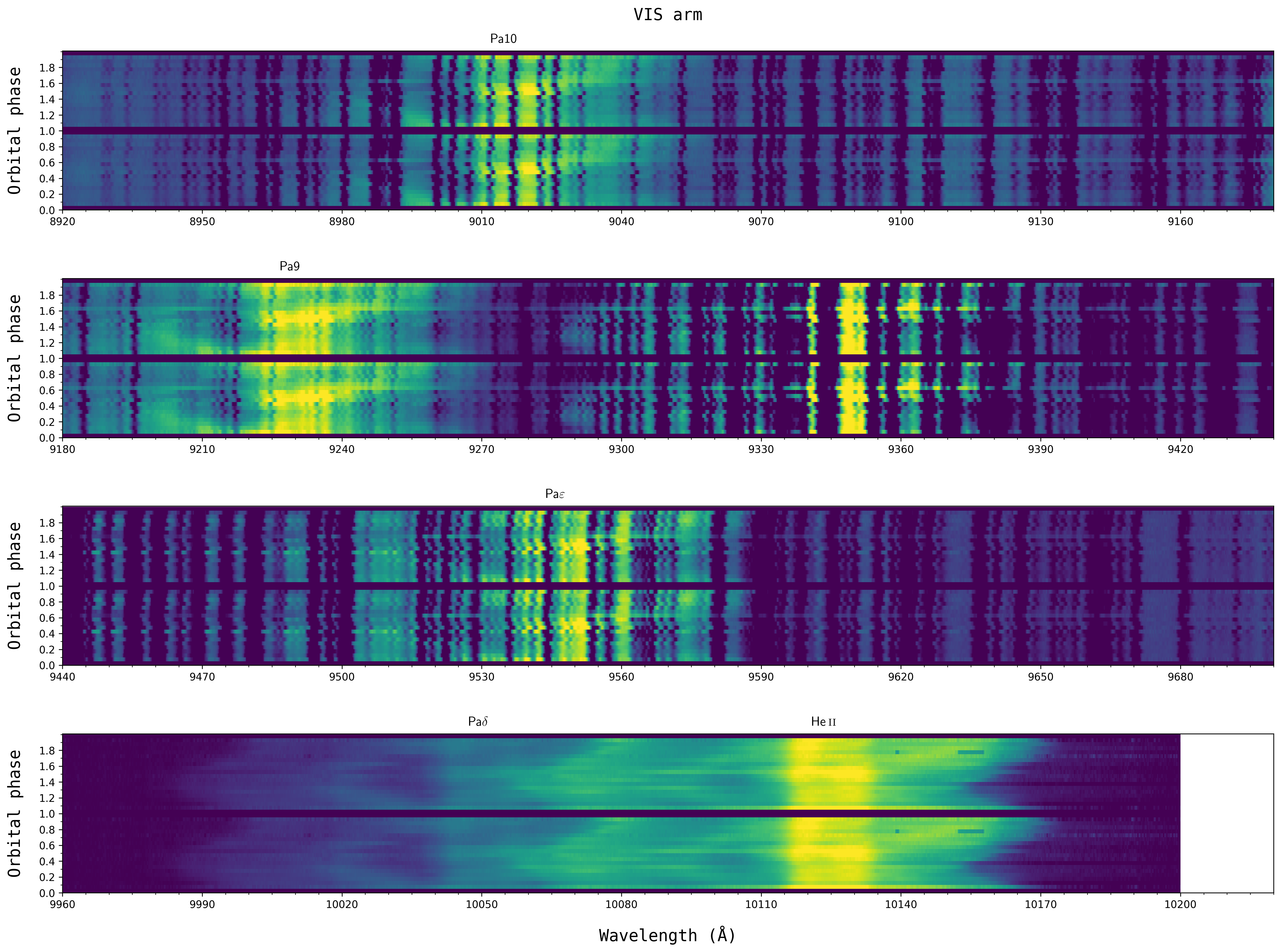}
\caption{}
\label{fig:trailed_VIS_04}
\end{center}
\end{figure}
\end{landscape}

\begin{landscape}
\begin{figure}
\begin{center}
\includegraphics[width=0.85\linewidth]{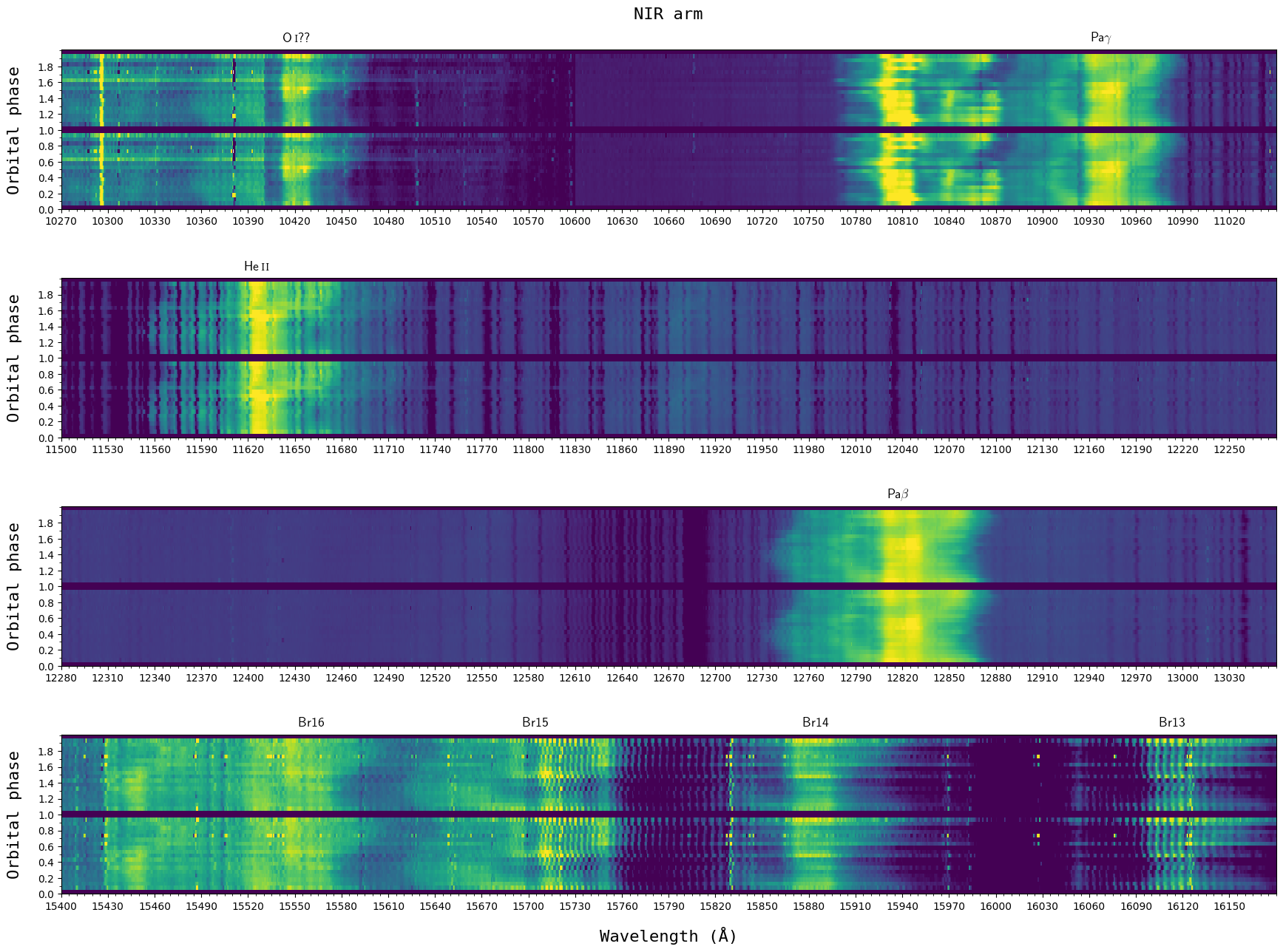}
\caption{X-Shooter NIR arm trailed spectra diagrams using all the data averaged into 20 orbital phase bins. The full cycle has been repeated once for clarity.}
\label{fig:trailed_NIR_01}
\end{center}
\end{figure}
\end{landscape}

\begin{landscape}
\renewcommand{\thefigure}{\thesection\arabic{figure} (cont.)}
\addtocounter{figure}{-1}
\begin{figure}
\begin{center}
\includegraphics[width=0.85\linewidth]{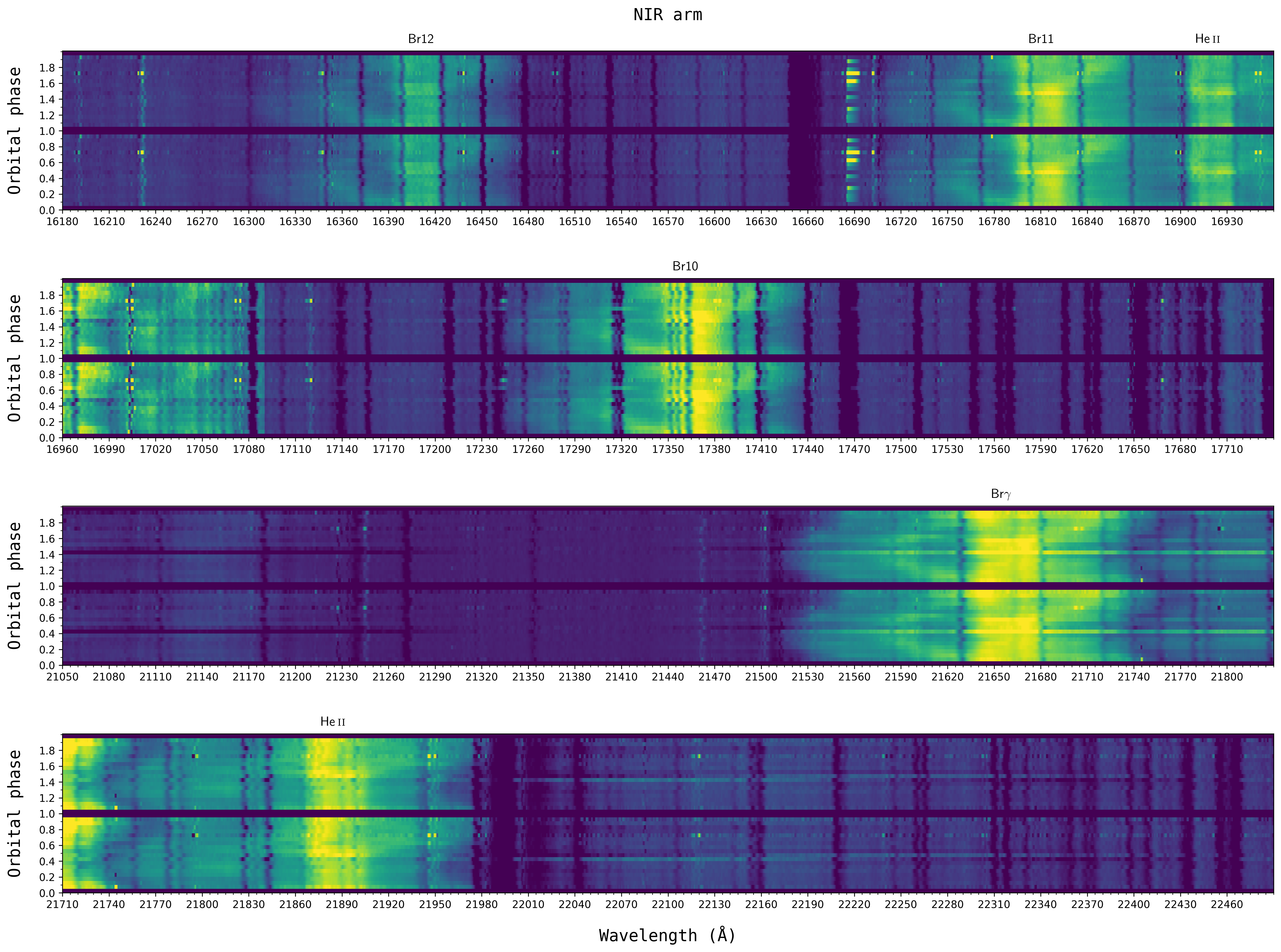}
\caption{}
\label{fig:trailed_NIR_02}
\end{center}
\end{figure}
\end{landscape}

\section{The Kinematic analysis of spectral line wings}\label{appendix:wings}

\begin{figure*}
\begin{center}
\includegraphics[width=0.49\linewidth]{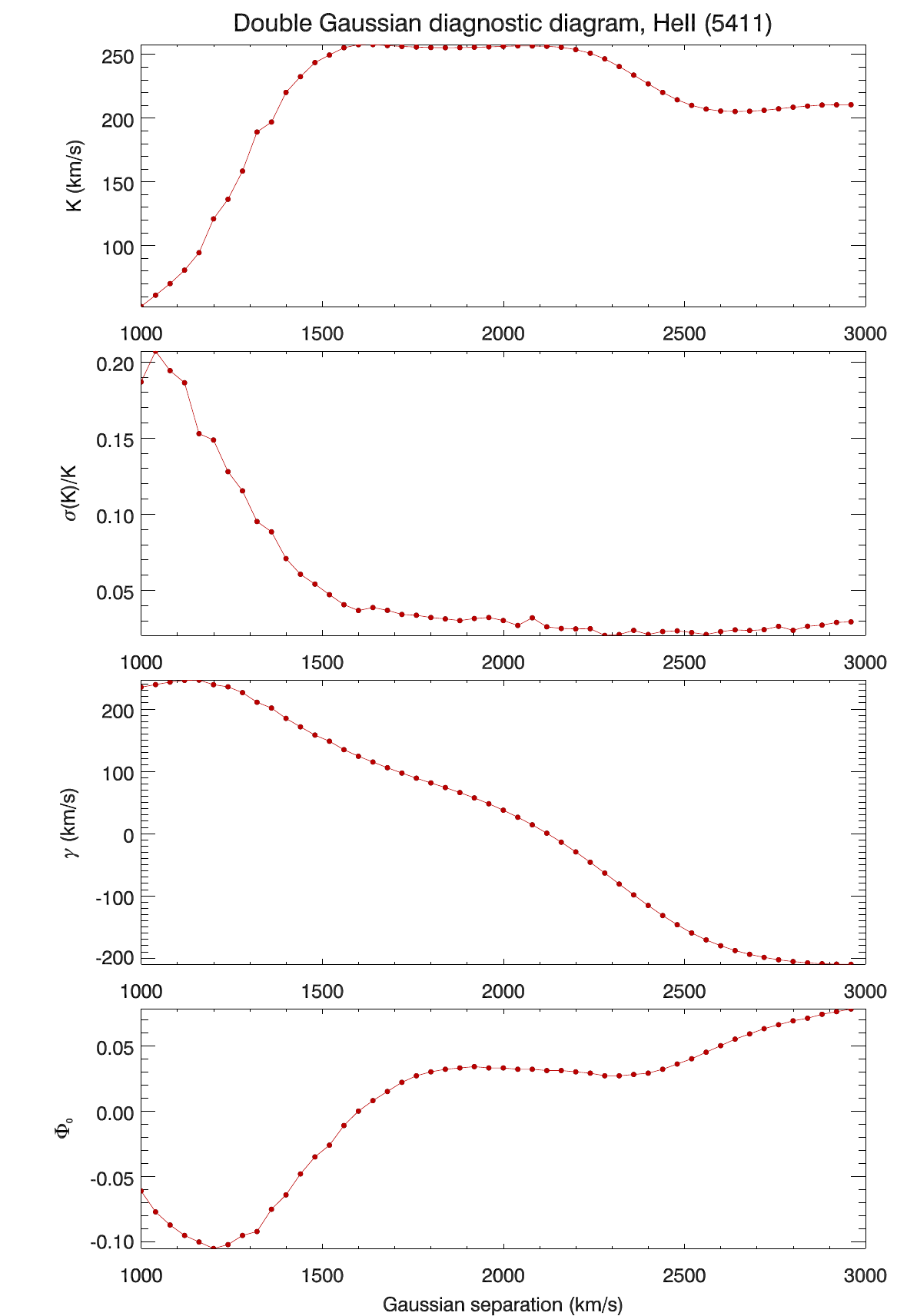}
\includegraphics[width=0.49\linewidth]{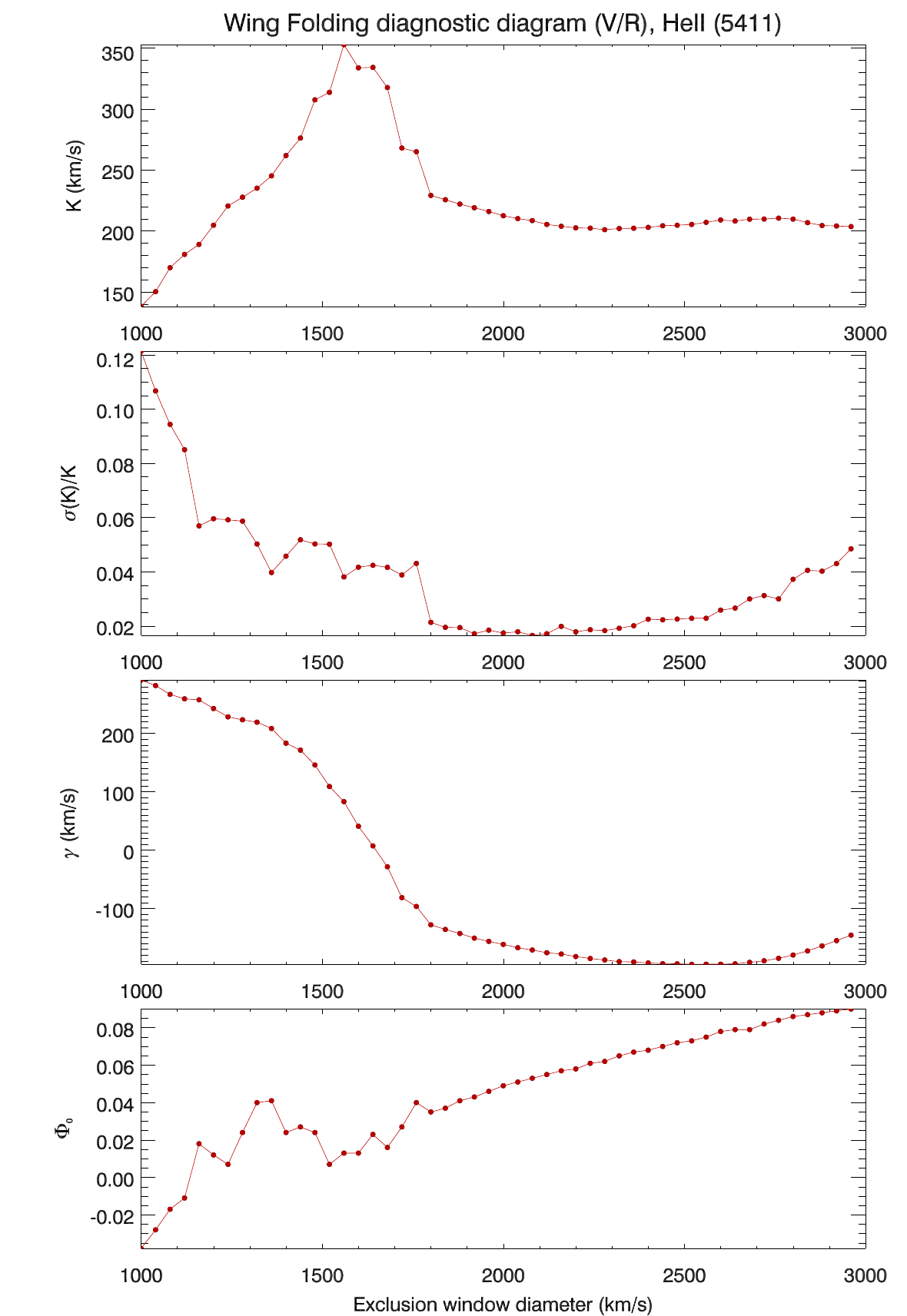}
\caption{
Comparison of diagnostic diagrams obtained using the double-Gaussian method (left) and the newly developed `wing-folding' technique (right).
}
\label{fig:dbldiag}
\end{center}
\end{figure*}

Since the emission-line profiles in \src\ are very complex and consist of multiple components, it is not trivial to extract the orbital velocities of the stellar components. In particular, we do not detect {\it any} spectroscopic signal that could be associated with the donor star. However, the very broad wings of the emission lines do appear to follow approximately the predicted movement of the primary (accreting) component, that is eclipsed at phase 0. In order to extract/measure this movement (i.e. determine the $K$ velocity of the primary), we initially utilised the double-Gaussian approach 
\citep{Schneider+Young1980, Shafter1983} with a Gaussian width of 250\;\kms, as described in Section~\ref{sec:broad:comp}.

As the results from the double-Gaussian analysis were not entirely conclusive (Fig.~\ref{fig:dbldiag}), we developed another method that we call `wing-folding'. In this technique, we apply an exclusion window (of variable extent) to the central part of the emission line profile and compare the relative flux of the line remaining in the blue and red wings for
different values of the line central velocity. The best radial velocity for each line profile is then determined by looking for a line-wing flux ratio $V/R = 1$. Applying this method for different widths of the central exclusion region produces a similar diagnostic diagram to the double-Gaussian method, but is {\it independent} of any functional form for the line wings. We show here the diagnostic diagrams using both methods (Fig.~\ref{fig:dbldiag}). We conclude from these that the observed $K$ velocity is very likely in the 200--250\,\kms\ range. Given that partially eclipses are seen in the light
curve, the inclination must be above $\approx\!65\degr$, i.e. the real $K$ has an upper limit of 220--275\;\kms. We also find that the $\gamma$ velocity of the central accretion disc, where we presume the high-velocity wing emission originates, is approximately --150\;\kms.

\bsp	
\label{lastpage}

\end{document}